%% file: WeightedGarblingMain.tex
\documentclass[12pt]{article}
\usepackage{amssymb}
\usepackage[nohead]{geometry}
\usepackage[singlespacing]{setspace}
\usepackage[colorlinks=false]{hyperref}
\usepackage{amsthm} 
\usepackage{amsmath}
\usepackage{natbib}

\usepackage[dvipsnames]{xcolor}
\usepackage{amsthm}
\usepackage{sgame}
\usepackage{hyperref}
\usepackage{cleveref}
\usepackage{verbatim}
\usepackage{titlesec}
\usepackage{subcaption}
\usepackage{thmtools}
\usepackage[T1]{fontenc}
\usepackage{enumitem}
\usepackage{subfiles}
\usepackage{lmodern}
\usepackage{graphicx}
\usepackage{pgfplots}
\usepgfplotslibrary{ternary}

  \theoremstyle{plain}
  
 \theoremstyle{definition}
  \newtheorem{example}{\protect\examplename}
  \theoremstyle{plain}
  \newtheorem{prop}{\protect\propositionname}
  \theoremstyle{plain}
  \newtheorem{cor}{\protect\corollaryname}
  \theoremstyle{plain}
  
  \theoremstyle{plain}
  \newtheorem{lem}{\protect\lemmaname}
\theoremstyle{plain}
\newtheorem{thm}{\protect\theoremname}

\theoremstyle{definition}
  \newtheorem{defn}{\protect\definitionname}

  \theoremstyle{remark}

  \theoremstyle{definition}

\DeclareMathOperator*{\supp}{supp}

\makeatother

  \providecommand{\assumptionname}{Assumption}
  \providecommand{\claimname}{Claim}
  \providecommand{\conjecturename}{Conjecture}
  \providecommand{\definitionname}{Definition}
  \providecommand{\examplename}{Example}
  \providecommand{\lemmaname}{Lemma}
  \providecommand{\propositionname}{Proposition}
\providecommand{\corollaryname}{Corollary}
\providecommand{\theoremname}{Theorem}

\renewcommand\thmcontinues[1]{Continued}

\begin{document}

\title{Weighted Garbling\thanks{We would like to thank Alex Bloedel, David Dillenberger, Jay Lu, Moritz Meyer-ter-Vehn, and Joe Ostroy for their comments. We also like to thank our audience at various seminars for their feedback, including the 2022 Asian Meeting of the Econometric Society in East and South-East Asia, the 2023 Kansas Workshop in Economic Theory, the 2023 European Winter Meeting of the Econometric Society, the 2024 North American Winter Meeting, and the 2025 Econometric Society World Congress. All remaining errors are ours.}}
\author{Daehyun Kim\thanks{Department of Economics, UCLA. E-mail: \href{mailto:dkim85@outlook.com}{\nolinkurl{dkim85@outlook.com}}}\and Ichiro Obara\thanks{Department of Economics, UCLA. E-mail: \href{mailto:iobara@econ.ucla.edu}{\nolinkurl{iobara@econ.ucla.edu}}}}

\date{March 7, 2026}
\maketitle

\singlespacing

\begin{abstract}
We introduce an information order on experiments based on \emph{weighted garbling}, a generalization of the standard notion of garbling. In this order, an experiment is more informative than another if the latter is a weighted garbling of the former. We show that this is equivalent to ordinary garbling conditional on a payoff-irrelevant event. We also characterize the order in terms of induced posterior belief distributions, showing that it depends only on their support.  
Our main results provide two decision-theoretic characterizations of this order. First, in static decision problems, one experiment dominates another if and only if its value of information is at least a fixed fraction of the other's across all problems. Second, in a class of stopping time problems with a hidden Markov process and repeated experimentation, one experiment dominates another if and only if it yields weakly higher expected payoffs for every problem with a regular prior.
\end{abstract}
\strut

\strut


\pagebreak


\pagebreak

\onehalfspacing

\section{Introduction}
\label{sec:1}

\subfile{section1}

\section{Weighted Garbling}
\label{sec:2}

\subfile{section2}

\section{Belief-Based Characterization of Weighted-Garbling Order}
\label{sec:3}

\subfile{section3}

\section{Weighted-Garbling Order and Payoff Guarantee}
\label{sec:4}

\subfile{section4}

\section{Weighted-Garbling Order and Optimal Stopping Problems with Hidden Markov Processes}
\label{sec:5}

\subfile{section5}

\section{Discussion: Weighted-Garbling Order and Other Extensions of Blackwell Order}
\label{sec:6}

\subfile{section6}

\section{Conclusion}
\label{sec:7}
\subfile{section7}

\newpage

\appendix
\section{Appendix}
\label{sec:app}

\subfile{appendix}

\bibliography{weighted_garbling}    
\bibliographystyle{chicago}

\end{document}

%% file: section1.tex
Consider an agent facing a decision problem in which her payoff depends on both her action and the (hidden) state of nature. Before choosing an action, she may conduct an experiment to obtain information about the state. A fundamental question in Economics and Statistics is when one piece of information should be regarded as more informative than another. We can regard one experiment (or information structure) as being more informative than another if the agent would achieve a weakly higher expected payoff with the former in every decision problem for another \cite[]{BBS_1949}. Blackwell's theorem shows that this definition is equivalent to the sufficiency of the former experiment for the latter experiment, i.e., the latter can be generated by garbling the former \cite[]{Blackwell_1951, Blackwell_1953_AMS}. Although Blackwell's information order provides a natural partial order over experiments, it is quite demanding, and many experiments are not comparable under it.

In this paper, we introduce a preorder, the \emph{weighted garbling order}, which is based on a relaxed notion of garbling and hence is more permissive than the Blackwell order. We show that it captures the idea that one experiment may be \emph{conditionally} more informative than another. This order admits a simpler characterization than Blackwell's in terms of posterior belief distributions, making it more practical for comparing information structures in real-world applications. Our main results provide various foundations for the weighted garbling order. First, we characterize it in terms of the worst-case ratio of the values of information generated by two experiments. Second, we show that an experiment is more informative in the weighted garbling sense if and only if it is more useful in a dynamic environment with changing states when the same experiment can be repeated many times before a decision is made.

Our starting point is the garbling characterization of the Blackwell order. An experiment $\Pi = (S, \{\pi_\theta\}_\theta)$ consists of a signal space $S$ and conditional distributions $\pi_\theta \in \Delta(S)$ for each $\theta \in \Theta$, where $\Theta$ denotes the set of states of nature. We say that $\Pi$ is a Blackwell garbling of $\Pi' = (S', \{\pi'_\theta\}_\theta)$ if there exists a Markov kernel $\phi: S' \to \Delta (S)$ such that $\pi_\theta (X) = \int_{S'} \phi(X|s') \pi'_\theta (ds')$ for every Borel subset $X \subseteq S$ and $\theta \in \Theta$. Intuitively, $\Pi$ is obtained from $\Pi'$ by adding noise that does not contain any additional information about the state of nature. We generalize this notion by allowing each signal $s' \in S'$ to be reweighted. Formally, $\Pi$ is a \textit{weighted garbling} of $\Pi'$ if there exist nonnegative weights $\gamma: S' \rightarrow \mathbb{R}_+$ and Markov kernel $\phi: S' \to \Delta (S)$ such that $\pi_\theta (X) = \int_{S'} \phi (X|s') \gamma_{s'} \pi'_\theta (ds')$ for every Borel subset $X \subseteq S$ and $\theta \in \Theta$. Weighted garbling reduces to Blackwell garbling when we can take $\gamma = 1$ as the weight. The infimum of $\sup_{S'} \gamma_{s'}$ over all weights that satisfy the above definition is called the \emph{size} of the weighted garbling relationship, which is always at least $1$ and provides a quantitative measure of how close a weighted garbling relationship is to a Blackwell garbling relationship.

There are several ways to characterize the weighted garbling order. Our first characterization is expressed in terms of posterior beliefs. Recall that an experiment $\Pi'$ is more informative than another experiment $\Pi$ in the Blackwell order if and only if, for every full-support prior, the posterior belief distribution induced by $\Pi'$ is a mean-preserving spread of the posterior belief distribution induced by $\Pi$. We show that $\Pi'$ is more informative than $\Pi$ in the weighted garbling order if and only if there exists a distribution over posterior beliefs that is ``close'' to the posterior belief distribution generated by $\Pi'$ and is a mean-preserving spread of the posterior belief distribution generated by $\Pi$. Furthermore, this ``closeness'' is tightly linked to the size of the weighted garbling. For experiments with finite signal spaces, this condition simplifies to a geometric requirement: the convex hull of posterior beliefs induced by $\Pi'$ contains that induced by $\Pi$. For instance, when there are only two states (so posterior beliefs are scalars), $\Pi'$ is more informative than $\Pi$ in the weighted garbling order if and only if the most extreme posterior beliefs generated by $\Pi'$ lie outside the corresponding extreme posterior beliefs generated by $\Pi$. This criterion is far easier to verify in practice than checking the full mean-preserving spread condition. In addition, the size of a weighted garbling, which quantifies the ``distance'' between a weighted garbling order and the Blackwell order, can be easily computed from data by solving a simple linear programming problem for experiments with finite signal spaces.

Our main results provide several decision-theoretic characterizations of the weighted garbling order. We begin by showing that weighted garbling is equivalent to \emph{conditional informativeness}. Specifically, experiment $\Pi'$ is conditionally more informative than $\Pi$ if there exists an (uninformative) event for $\Pi'$ such that, conditional on that event, $\Pi'$ becomes Blackwell more informative than $\Pi$. The maximum probability of such an event is shown to be exactly the reciprocal of the size of the weighted garbling relationship. 
A related characterization links weighted garbling to the worst-case performance of $\Pi'$ relative to $\Pi$ (\autoref{thm:n1}). More precisely, $\Pi'$ is more informative than $\Pi$ in the weighted garbling order if and only if there is a strictly positive lower bound for the ratios of the marginal value of information for $\Pi'$ to the marginal value of information for $\Pi$ uniformly across all decision problems.\footnote{For a decision problem $\mathcal{A} = (A, u, \mu_0)$, which consists of actions, payoff, and prior belief, the marginal value of information for $\Pi$ is $V^{\mathcal{A}}(\Pi) - V^{\mathcal{A}}(\varnothing)$, where $V^{\mathcal{A}}(\Pi)$ is the optimal expected payoff for $\mathcal{A}$ under $\Pi$ and $V^{\mathcal{A}}(\varnothing)$ is the optimal expected payoff for $\mathcal{A}$ with no experiment.} Furthermore, as suggested by the above conditional-informativeness interpretation, the maximum tight lower bound corresponds exactly to the reciprocal of the size of the weighted garbling. For example, $\Pi$ is a weighted garbling of $\Pi'$ of size $0.5$ if and only if the marginal value of $\Pi'$ is guaranteed to be at least $50\%$ of the marginal value of $\Pi$ for every decision problem.

For a completely different characterization (\autoref{thm:4}), we study a class of optimal stopping problems in which the state evolves according to a Markov process and the decision maker may repeatedly conduct the same experiment before making a single irreversible choice by some terminal date $T$.\footnote{\cite{Monahan_1980_OR} analyzes a related optimal stopping problem with evolving states and costly information acquisition. This framework can also be viewed as a variant of the ``secretary problem,'' where applicants arrive sequentially, and a patient administrator aims to hire a high-quality candidate using one of two sets of interview questions. Our main result implies that if one set of questions is superior to another in the weighted garbling order, then the former yields a higher expected quality of the eventual hire when the applicant pool is sufficiently large.} This dynamic environment essentially constitutes an optimal stopping problem with a hidden Markov process. Within this class of problems, we characterize the weighted garbling order of experiments through the decision maker's optimal expected payoff. The basic idea is that a more informative experiment in terms of weighted garbling is more useful when the decision maker can conduct the same experiment many times before making a decision. Our main result (\autoref{thm:4}) shows that if an experiment is more informative than another in the weighted garbling order, then we can find the number of periods $T'$ such that every decision maker achieves a weakly higher expected payoff with the former experiment in every dynamic decision problem with any terminal date $T \geq T'$, provided the initial belief is not too ``extreme'' and transient. 
Conversely, if an experiment is not more informative than another in the weighted garbling order, then there exists a dynamic decision problem in this class (with the initial belief satisfying the same regularity condition) in which the decision maker obtains a strictly higher expected payoff with the latter experiment given any large $T$.

\vspace{3mm}

\textbf{Related Literature}

\vspace{2mm}

This paper contributes to the literature on the comparison of experiments/information structures, pioneered by Blackwell \cite[]{Blackwell_1951, Blackwell_1953_AMS}. Blackwell shows that one experiment is sufficient for another if and only if it is more informative in the sense of yielding a weakly higher expected payoff in every decision problem \cite[]{BBS_1949}. Because this requirement demands dominance across all decision problems, many experiments are not comparable under the Blackwell order. A large body of work studies more permissive information orders by restricting attention to particular classes of decision problems \cite[]{Lehmann_1988_AS, Persico_2000_ECMA, CGS_2013_AER, AL_2018}.

The idea of capturing informativeness through the dispersion of posterior beliefs is common. The characterization based on the mean-preserving spread condition already appears in \cite[]{Blackwell_1951, Blackwell_1953_AMS}.\footnote{However, Blackwell's results are not stated in Bayesian language. See \cite{Kihlstrom_1984} for a Bayesian exposition of the results.} When the state is one dimensional, \cite{GP_2010_ECMA} introduce an information order based on the second-order stochastic dominance of posterior mean distributions to study an auctioneer's incentives to disclose information. This information order is weaker than Blackwell's order and prior-dependent. For finite experiments, the weighted garbling order reduces to the inclusion order over the convex hulls of posterior beliefs. \cite{Wu_2023_EL} shows that this inclusion order coincides with the Blackwell order when the more dispersed posteriors satisfy affine independence. \cite{JO_1984_RES} introduce a star-shaped spreading of beliefs, which is a stronger requirement than the mean-preserving spread condition, and show that this particular type of dispersion of posteriors enhances the relative value of perfectly flexible actions.

There is a line of research that examines decision makers who may freely observe an arbitrary number of i.i.d. samples about a hidden state before acting \cite[]{MS_2002_ECMA, Azrieli_2014_ECMA, MPST_2021_ECMA, MS_2002_ECMA}. \cite{MS_2002_ECMA} define one experiment to be more informative than another if, for every decision problem, there exists an integer $N$ such that for all $n \geq N$, the decision maker achieves a higher expected payoff from the former experiment with $n$ samples. \cite{Azrieli_2014_ECMA} proposes a stronger notion requiring this threshold integer to be uniform across all decision problems and shows that the two criteria are not equivalent. \cite{MPST_2021_ECMA} further demonstrate that, in binary-state environments, this stronger notion is characterized by larger R\'{e}nyi divergences. In contrast to these papers, we study a dynamic environment with an evolving state. We discuss how these information orders relate to the weighted garbling order in \Cref{sec:6}.

The idea of weighted garbling originates in our earlier work \cite[]{KO_2023} on stochastic games with imperfect monitoring, where we introduced a version of weighted garbling in a specific setting. There, we showed that the limit equilibrium payoff set expands whenever the monitoring structure improves in the sense of weighted garbling. In the present paper, we develop this notion more fully and study weighted garbling in a general decision-theoretic framework.

The remainder of the paper is organized as follows. \Cref{sec:2} introduces statistical decision problems and the basic definitions, including the key notions of conditional informativeness and weighted garbling. \Cref{sec:3} develops a belief-based characterization of the weighted garbling order. \Cref{sec:4} examines weighted garbling in statistical decision problems and presents one of our main results, which links weighted garbling to guaranteeing a decision maker's expected payoff from a more informative experiment relative to that from a less informative one. \Cref{sec:5} provides another main result, characterizing weighted garbling through the value of information in a class of dynamic decision problems. \Cref{sec:6} discusses the relationship between the weighted garbling order and other extensions of Blackwell's order. \Cref{sec:7} concludes.

%% file: section2.tex
\subsection{Decision Problems and Experiments}

Let $\Theta$ be a finite set of states of nature. Given $\Theta$, a \emph{decision problem} is defined as a triple $\mathcal{A} = (A, u, \mu_0)$, where $A$ is the set of actions, $u: A \times \Theta \to \mathbb{R}$ is the payoff function, and $\mu_0 \in \Delta(\Theta)$ is the prior belief for a decision maker (DM).\footnote{It is assumed that every space is Polish and every function is measurable. For any polish space $X$, $\Delta(X)$ is the set of Borel probability measures on $(X, \mathcal{B}_X)$, where $\mathcal{B}_X$ is the Borel $\sigma$-field over $X$.}   

An \emph{experiment} is defined as $\Pi = (S, \{\pi_\theta\}_{\theta \in \Theta})$, where $S$ is a signal space and $\pi_\theta \in \Delta(S)$ is a probability measure on $S$ for each $\theta$. $\Pi$ is a \emph{finite experiment} if $S$ is a finite set.

Given a decision problem $\mathcal{A}$ and an experiment $\Pi$, the DM chooses an action after observing a signal generated by $\Pi$. The DM's strategy is a Markov kernel $\sigma: S \rightarrow 
\Delta(A)$.\footnote{Recall that for any Markov kernel $f:X \rightarrow \Delta(Y)$, $f( \cdot | x)$ is a probability measure on $Y$ for each $x \in X$ and, for each Borel set $B \subseteq Y$, $f(B | \cdot)$ is a measurable function from $X$ to $\mathbb{R}_+$.} Let $U^{\mathcal{A}}(\sigma ; \Pi) = \sum_{\theta \in \Theta} \int_{S} \int_{A}u(a, \theta)\sigma(da|s) \pi_\theta(ds) \mu_0(\theta)$ be the ex-ante expected payoff from strategy $\sigma$. Then $V^{\mathcal{A}}(\Pi) = \sup_{\sigma} U^{\mathcal{A}}(\sigma ; \Pi)$ represents the supremum of the expected payoffs the DM can achieve for decision problem $\mathcal{A}$ given experiment $\Pi$.

\subsection{Weighted Garbling}

An experiment $\Pi =(S, \{\pi_\theta \}_{ \theta \in\Theta} )$ is a \emph{Blackwell garbling} of another experiment $\Pi^\prime =(S^\prime, \{\pi^\prime_\theta\}_{ \theta \in \Theta} )$ if there exists a Markov kernel $\phi: S' \rightarrow \Delta(S)$ that satisfies
$$\pi_\theta(X) = \int_{S'} \phi (X|s')\pi^\prime_\theta(ds'),  \quad \forall \theta \in \Theta.
$$
for every Borel set $X \subseteq S$.

We relax the notion of Blackwell garbling by allowing for flexible weights over signals. For experiment $\Pi' = (S', \{\pi'_\theta\}_\theta)$, let $\Gamma_{S'}$ be the set of all nonnegative bounded functions on $S'$.

\begin{defn}[\textbf{Weighted Garbling and Size}]
\label{defn:WG}
An experiment $\Pi = (S, \{\pi_\theta\}_{\theta \in \Theta})$ is a \emph{weighted garbling} of experiment $\Pi^\prime = (S', \{\pi'_\theta\}_{\theta \in \Theta})$ if there exist $\gamma \in \Gamma_{S'}$ and a Markov kernel $\phi: S' \rightarrow \Delta(S) $ that satisfies
\begin{equation}    
\label{eq:nnn1}                                      
\pi_\theta(X) = \int_{S^\prime}\gamma_{s^\prime} \phi(X |s^\prime)\pi^\prime_\theta(ds^\prime) \quad \forall \theta \in \Theta
\end{equation}   
for every Borel set $X \subseteq S$.\footnote{Equivalently, we can define weighted garbling by $\pi_\theta(X) = \int_{S^\prime} \Phi(X |s^\prime)\pi^\prime_\theta(ds^\prime)$, where $\Phi(\cdot| s^\prime)$ is a (uniformly) finite positive measure on $(S, \mathcal{B}_S)$. In fact, if we define $\gamma: s^\prime \rightarrowtail \Phi(S|s^\prime)$, which is measurable, and normalize $\Phi$ by $\gamma$ at each $s'$, then we obtain a probability measure $\phi(\cdot|s^\prime) = \frac{\Phi(\cdot|s^\prime)}{\gamma_{s^\prime}}$ for each $s^\prime$.}

$\Pi$ is a weighted garbling of $\Pi'$ with size $\beta \in [1, \infty)$ if $\beta$ is the infimum of $\sup_{s' \in S'} \gamma_{s'}$ over all $\gamma \in \Gamma_{S'}$ that satisfies \eqref{eq:nnn1}.
\end{defn}       

We call $\gamma$ \emph{weight function}. When $\Pi$ is a weighted garbling of $\Pi'$ with some weight $\gamma$, we say $\Pi^\prime$ is WG-more informative than $\Pi$ and denote $\Pi' \succeq_{WG} \Pi$. For any weight $\gamma \in \Gamma_{S'}$, let $\overline{\gamma} = ess\sup \ \gamma < \infty$ be the essential supremum of $\gamma$ on $S'$ with respect to a uniform (or any full-support) prior on $\Theta$.\footnote{So, $\overline{\gamma}$ is the infimum of $a \geq 0$ such that $\gamma_{s'} > a$ is with $0$ probability at every state.} We assume $\overline{\gamma}$ is equal to $ \sup_{s' \in S'} \gamma_{s'}$ without loss of generality.

There are many useful ways to interpret weighted garbling.\footnote{We discussed these interpretations in \cite{KO_2023}.} For one interpretation, observe that $\gamma \pi'_\theta$, which is defined by $\gamma \pi^\prime_\theta(X^\prime) = \int_{X^\prime} \gamma_{s^\prime} \pi^\prime_\theta(ds')$ for any Borel set $X'$ of $S'$, is a probability measure on $S'$ as the expected value of $\gamma$ is $1$ for each $\theta \in \Theta$. This implies that each weight $\gamma$ generates another experiment $(S^\prime, \{\gamma \pi^\prime_\theta\}_{\theta \in \Theta})$ from $\Pi^\prime$, which we denote by $\gamma \Pi^\prime$. Intuitively, we can interpret weighted garbling as a two-step transformation. Weight $\gamma$ transforms an experiment $\Pi'$ to another experiment $\gamma \Pi'$ by distorting the signal distributions in a state-independent way. Then $\Pi$ is obtained as a standard Blackwell garbling of this experiment $\gamma \Pi'$. We elaborate on this viewpoint more fully in the next section.

\begin{example}\label{exmp:1}

Suppose there are two possible states, $\Theta = \left\{\theta_1, \theta_2\right\}$. Let $\Pi = \left(S, \{\pi_\theta\}_{\theta \in \Theta}\right)$ be an experiment with two signals $S = \left\{s_1, s_2\right\}$ and conditional distributions such that $\pi_{\theta_1}(s_1) =\pi_{\theta_2}(s_2) = q > 0.5$. Let $\Pi^\prime = \left(S^\prime, \{\pi^\prime_\theta\}_{\theta  \in \Theta}\right)$ be another experiment with three signals $S^\prime = \left\{s_0^\prime, s_1^\prime, s_2^\prime\right\}$ and the following signal distributions: $\pi^\prime_{ \theta_i} (s_0^\prime) = 0.5$ for $i=1,2$ and $\pi'_{ \theta_1} (s_1^\prime) = \pi'_{ \theta_2} (s_2^\prime) = 0.5 q^\prime$, where $q^\prime > q$ (see \Autoref{fig:n1}). Note that $s_0'$ is essentially a null signal that does not convey any information about the state. 

\begin{figure}
	
	\centering
	\begin{tabular}{|c|c|c|}
		\hline
		$\pi_\theta$    & $\theta_1$ & $\theta_2$ \\ \hline
		$s_1$ & $q$        & $1-q$      \\ \hline
		$s_2$ & $1-q$      & $q$        \\ \hline
	\end{tabular}
	\quad
	\begin{tabular}{|c|c|c|}
		\hline
		$\pi'_\theta$    & $\theta_1$   & $\theta_2$   \\ \hline
		$s_1'$ & $0.5 q'$     & $0.5 (1-q')$ \\ \hline
		$s_2'$ & $0.5 (1-q')$ & $0.5 q'$     \\ \hline
		$s_0'$ & $0.5$        & $0.5$        \\ \hline
	\end{tabular}
	\caption{Experiments in Example 1. $q' > q$}
	\label{fig:n1}
\end{figure}

In this example, $\Pi$ is a weighted garbling of $\Pi^\prime$ with respect to weight $\gamma_{s_0} =0, \gamma_{s_1'} = \gamma_{s_2'} =2$ and $\phi$ such that $\phi(s_i|s_i') = \frac{q'+q-1}{2q'-1}$ for $i=1,2$. For this weight, $\overline{\gamma} =2$. However, $\Pi$ is a weighted garbling of $\Pi^\prime$ with size less than $2$, since we can reduce the weight of the informative signals further while increasing the weight of the null signal. Specifically, $\Pi$ is also a weighted garbling of $\Pi^\prime$ with respect to weight $\gamma_{s_0^\prime} = \max \left \{\frac{2(q'-q)}{q'-0.5}, 1 \right\}$, $\gamma_{s_i^\prime} = 2 - \gamma_{s_0^\prime}$ for $i=1,2$ and $\phi$ such that $\phi(s_i|s^\prime_i) =1$ for $i=1,2$ and $\phi(s_i|s^\prime_0) =0.5$. In fact, the size of this weighted garbling is $\beta = 2 - \gamma_{s_0^\prime}$.

\end{example}

In the above example, we see that $\Pi^\prime$ is ``conditionally more informative'' than $\Pi$ in the sense that it is more informative when either $s^\prime_1$ or $s^\prime_2$ realizes. In \Cref{sec:4}, we show that weighted garbling is indeed equivalent to a notion of conditional informativeness. 

\begin{example}\label{exmp:2}
Assume there are two states $\theta_1 \neq \theta_2 \in \mathbb{R}$ on the real line. Experiment $\Pi = (S, \{ \pi_\theta\}_{\theta \in \Theta})$ generates a signal equal to the state plus a normal noise with variance $1$. Thus, $\pi_\theta$ is the normal distribution with mean $\theta$ and variance $1$ for each $\theta = \theta_1, \theta_2$. 

For $\Pi^\prime = (S', \{ \pi'_\theta \}_{ \theta \in \Theta})$, suppose the signal is the state plus a normal noise with a smaller variance $\sigma^2 < 1$. However, the signal realization $s'$ is unobservable with probability $50\%$ if and only if $s'$ lies outside a given interval $[-\overline{s}', \overline{s}']$ for some $\overline{s}' >0$. Formally, $S^\prime  = \mathbb{R} \cup \{\emptyset\}$, where $\emptyset$ represents observing nothing.

Then $\Pi$ is a weighted garbling of $\Pi^\prime$, since we can use the following weight function: $\gamma_{s'}=1$ for $s' \in [-\overline{s}', \overline{s}']$, $\gamma_{s'} =2$ for $s' \notin [-\overline{s}', \overline{s}']$, and $\gamma_{\emptyset} = 0$.

\end{example}

\subsection{Transitivity of Weighted Garbling Order}
\label{subsec:2.4}

We briefly examine some transitive property of the weighted garbling order. 

\begin{prop}
	\label{prop:n2}
	If $\Pi$ is a weighted garbling of $\Pi'$ with weight $\gamma_1$ and $\Pi'$ is a weighted garbling of $\Pi''$ with weight $\gamma_2$, then $\Pi$ is a weighted garbling of $\Pi''$ with some weight $\gamma$ such that $\overline{\gamma} \leq \overline{\gamma}_1 \overline{\gamma}_2$.
\end{prop}
\begin{proof}
	See \Autoref{sec:app}.
\end{proof}

So, the weighted garbling order satisfies transitivity. Note that this transitivity holds for the sizes as well. The above result implies that $\Pi$ is a weighted garbling of $\Pi'$ with size $\beta$ less than or equal to $\overline{\gamma}_1 \overline{\gamma}_2$ for any weight $\gamma_1$ and $\gamma_2$. Taking the infimum of $\overline{\gamma}_1$ and $\overline{\gamma}_2$, we can obtain $\beta \leq \beta_1 \beta_2$. On the other hand, the weighted garbling order with respect to a given size is not transitive. This proposition shows that a particular type of transitivity holds once the sizes of weighted garblings are taken into account.

%% file: section3.tex
In this section, we provide a characterization of weighted garbling in terms of the distribution of posterior beliefs, which is used to prove our main results and is also of independent interest. For Blackwell garbling, it is known that $\Pi$ is a garbling of $\Pi^\prime$ if and only if the posterior belief for the latter is a mean-preserving spread of the posterior belief for the former \cite[]{Blackwell_1953_AMS}. We show that weighted garbling corresponds to a simple relaxation of the mean-preserving spread condition.

Given any prior distribution $\mu_0 \in \Delta (\Theta)$ and experiment $\Pi$, let $Q^{\Pi}_{\mu_0} \in \Delta(\Delta(\Theta))$ be the distribution of posterior belief induced by $\Pi$.\footnote{Given $\mu_0$, there is a measurable mapping that maps each signal outcome $s$ to a posterior belief. Let $\pi^*_\theta$ be the pushforward measure on $\Delta(\Theta)$ of $\pi_\theta$ by this mapping. Then $Q^{\Pi}_{\mu_0} = \sum_{\theta \in \Theta} \mu_0(\theta) \pi_\theta^* $.}
	
$P \in \Delta(\Delta(\Theta))$ is a \emph{mean-preserving spread} of $Q \in \Delta(\Delta(\Theta))$ if there exists a Markov kernel $f: \Delta(\Theta) \rightarrow \Delta(\Delta(\Theta))$ such that 
\begin{itemize}
	\item $P(X) = \int_{\Delta(\Theta)} f(X|q)Q(dq)$ for any Borel set $X \subseteq \Delta(\Theta)$.
	\item $v = \int_{\Delta(\Theta)} q f(dq|v)$ for $Q$-almost all $v \in \Delta(\Theta)$.
\end{itemize}

Our theorem shows that weighted garbling is characterized by replacing the condition of $Q^{\Pi^\prime}_{\mu_0}$ being a mean-preserving spread of $Q^{\Pi}_{\mu_0}$ by the existence of a mean-preserving spread $Q$ that satisfies some strong form of absolute continuity with respect to $Q^{\Pi^\prime}_{\mu_0}$.

\begin{thm}
\label{thm:1}
Let $\Pi = (S, \{\pi_\theta \}_{ \theta \in\Theta} )$ and $\Pi' = ( S', \{\pi'_\theta \}_{ \theta \in\Theta} )$ be experiments. 
\begin{enumerate}
\item If $\Pi$ is a weighted garbling of $\Pi'$ with weight $\gamma \in \Gamma_{S'}$, then for every full-support prior $\mu_0 \in \Delta(\Theta)$, there exists $Q \in \Delta\left(\Delta(\Theta)\right)$ such that $Q$ is a mean-preserving spread of $Q^{\Pi}_{\mu_0}$ and satisfies $\sup_{X \in \mathcal{B}_{\Delta(\Theta)}} \frac{Q(X)}{Q^{\Pi'}_{\mu_0}(X)} = \overline{\gamma}$.\footnote{Even when $\mu_0$ is not a full-support prior, the same equality still holds if $ \overline{\gamma}$ is replaced with the essential supremum with respect to $\sum_{\theta}\mu_0(\theta) \pi_\theta$.}

\item If there exists a full-support prior $\mu_0 \in \Delta(\Theta)$ and $Q \in \Delta\left(\Delta(\Theta)\right)$ such that $Q$ is a mean-preserving spread of $Q^{\Pi}_{\mu_0}$ and satisfies $\sup_{X \in \mathcal{B}_{\Delta(\Theta)}} \frac{Q(X)}{Q^{\Pi'}_{\mu_0}(X)}< \infty$, then $\Pi$ is a weighted garbling of $\Pi'$ with some weight $\gamma \in \Gamma_{S'}$ such that $\overline{\gamma} = \sup_{X \in \mathcal{B}_{\Delta(\Theta)}} \frac{Q(X)}{Q^{\Pi'}_{\mu_0}(X)}$.
\end{enumerate}
\end{thm}

Note that $\sup_{X \in \mathcal{B}_{\Delta(\Theta)}} \frac{Q(X)}{Q^{\Pi'}_{\mu_0}(X)} = \overline{\gamma}$ provides a detailed information about the gap between $Q$ and $Q^{\Pi^\prime}_{\mu_0}$. This condition implies that both $\frac{Q(X)}{Q^{\Pi'}_{\mu_0}(X)}$ and $\frac{1- Q(X)}{1 - Q^{\Pi'}_{\mu_0}(X)} = \frac{Q(X^c)}{Q^{\Pi'}_{\mu_0}(X^c)}$ are at most $\overline{\gamma}$. In particular, $Q$ and $Q^{\Pi'}_{\mu_0}$ coincide when $\overline{\gamma} =1$, i.e. $\Pi$ is a Blackwell garbling of $\Pi'$.

It is useful to consider the case of finite signal space to understand the theorem better and obtain a sharper result. In this case, $\sup_{X \in \mathcal{B}_{\Delta(\Theta)}} \frac{Q(X)}{Q^{\Pi'}_{\mu_0}(X)} < \infty$ means that the support of $Q$ is a subset of the support of $Q^{\Pi^\prime}_{\mu_0}$, which we denote by $\supp(Q^{\Pi'}_{\mu_0})$. Hence $Q$ can assign positive probability on and only on the set of posterior beliefs that can realize with $\Pi^\prime$. Then we can find $Q$ that satisfies the mean-preserving spread condition (hence $\Pi$ is a weighted garbling of $\Pi'$) if and only if every posterior belief in $\supp(Q^{\Pi}_{\mu_0})$ can be represented as a convex combination of posterior beliefs in $\supp(Q^{\Pi'}_{\mu_0})$.\footnote{\cite{Wu_2023_EL} shows that this condition becomes equivalent to Blackwell informativeness if the posterior beliefs for $\Pi'$ satisfy the condition he calls \emph{affine-independence}.} Hence, the following theorem follows from the above theorem for the finite case.

\begin{thm}\label{thm:2}
Consider finite experiment $\Pi$ and $\Pi'$. 
\begin{enumerate}
\item If $\Pi$ is a weighted garbling of $\Pi'$, then for every prior $\mu_0 \in \Delta(\Theta)$,  $\supp(Q^{\Pi}_{\mu_0}) \subset co\left(\supp(Q^{\Pi'}_{\mu_0})\right)$. 
\item If $\supp(Q^{\Pi}_{\mu_0}) \subset co\left(\supp(Q^{\Pi'}_{\mu_0})\right)$ for some full-support prior $\mu_0 \in \Delta(\Theta)$, then $\Pi$ is a weighted garbling of $\Pi'$.
\end{enumerate}
\end{thm}

\begin{proof}
	See \Autoref{sec:app}.
\end{proof}

This result suggests that it is empirically much easier to verify a weighted garbling relationship than to verify a garbling relationship for an arbitrary pair of experiments. When we compare two experiments in terms of weighted garbling, we only need to plot the posterior beliefs for each experiment and check whether the convex hull of one experiment's posteriors is nested in that of the other or vice versa. This contrasts with the Blackwell order, which involves the mean-preserving spread equalities. These would be very difficult to verify without some prior knowledge of the underlying information structures. 

We also like to note that, for the case of finite experiments, we can compute the size of weighted garbling explicitly by solving a linear programming problem. Suppose that each posterior belief $q$ for $\Pi$ can be expressed as a convex combination of posterior beliefs, hence $\Pi$ is a weighted garbling of $\Pi'$ by \autoref{thm:2}. For any possible convex combination $f(\cdot| q)$ for each $q$ (which satisfies $q = \sum_{q' \in \supp(Q^{\Pi'}_{\mu_0})} q' f(q'|q)$), we can derive $Q$ in \autoref{thm:1} on the support of $Q^{\Pi'}_{\mu_0}$ via $Q(q') = \sum_{q \in \supp(Q^{\Pi}_{\mu_0})} f(q'|q) Q^{\Pi}_{\mu_0}(q)$. Each such set of coefficients is implicitly associated with some weight $\gamma$ and $\overline{\gamma}$ is equal to $\max_{q' \in \supp(Q^{\Pi'}_{\mu_0})} \frac{Q(q')}{Q^{\Pi'}_{\mu_0}(q')}$. Then the problem of finding $\beta$ is equivalent to a linear programming problem to minimize $\max_{q'}\frac{Q(q')}{Q^{\Pi'}_{\mu_0}(q')}$ over all such coefficients $f(q'|q), \forall q' \in \supp(Q^{\Pi'}_{\mu_0}), \forall q \in \supp(Q^{\Pi}_{\mu_0})$ for the convex combinations.\footnote{We can use any full-support prior to do this. The solution is independent of the choice of full-support prior $\mu_0$.}

\autoref{thm:1} shows that an experiment is WG-more informative than another experiment if and only if there exists an experiment that satisfies some form of absolutely continuity with respect to the former experiment and is more Blackwell informative than the latter experiment. Hence, we can immediately obtain a version of another well-known characterization in terms of the convex order for weighted garbling. Remember that $P \in \Delta(\Delta(\Theta))$ dominates $Q \in \Delta(\Delta(\Theta))$ in the convex order when $\int_{\Delta(\Theta)} f(q) P(dq) \geq \int_{\Delta(\Theta)} f(q) Q(dq)$ for any continuous and convex function $f$.

\begin{thm}\label{thm:3}
	Let $\Pi$ and $\Pi'$ be experiments. 
	\begin{enumerate}
		\item If $\Pi$ is a weighted garbling of $\Pi'$, then for every prior $\mu_0 \in \Delta(\Theta)$, there exists $Q \ll Q^{\Pi^\prime}_{\mu_0}$ such that $Q$ dominates $Q^{\Pi}_{\mu_0}$ in the convex order. 
		
		\item If there exists a full-support prior $\mu_0 \in \Delta(\Theta)$ and $Q \ll Q^{\Pi^\prime}_{\mu_0}$ such that $Q$ dominates $Q^{\Pi}_{\mu_0}$ in the convex order, then $\Pi$ is a weighted garbling of $\Pi'$.
	\end{enumerate}
\end{thm}

%% file: section4.tex
\subsection{Conditional Informativeness}
\label{subsec:4.1}
In this section and the next section, we provide a foundation for the weighted garbling order in relation to decision problems. According to the standard Blackwell order, $\Pi$ is a garbling of $\Pi'$ if and only if $\Pi'$ is more informative than $\Pi$, i.e. $V^{ \mathcal{A}} (\Pi') \geq V^{ \mathcal{A}} (\Pi)$ for any decision problem $\mathcal{A}$ \cite[]{Blackwell_1951, Blackwell_1953_AMS}. This is a strong criterion that does not hold for many pairs of experiments. 

Since the weighted garbling order is more permissive than the garbling order, it corresponds to a weaker criterion in relation to decision problems. Our approach in this section is to apply a weaker criterion for the value of information.\footnote{Another approach is to examine the value of information only for some meaningful subclass of decision problems. For instance, \cite{CGS_2013_AER} relates the order induced from an entropy measure to a class of investment problems. 
} In this subsection, we observe that the weighted garbling order captures a notion of conditional informativeness in the sense that a WG-more informative experiment is more informative given some conditioning event that occurs with the same probability for all decision problems.

For any experiment $\Pi = (S, \{\pi_\theta\}_{\theta \in \Theta})$ and a measurable function $\eta: S \rightarrow [0,1]$, let $\pi^\eta_\theta$ be the probability measure on $S \times \{1,0\}$ that is defined by $\pi^\eta_\theta(X \times \{1\}) =\int_X \eta_s \pi_\theta(ds)$ for each Borel set $X \subseteq S$. So, $\eta$ represents a (regular) conditional probability over $\{0,1\}$ given each $s \in S$ that generates a joint distribution $\pi^\eta_\theta$ on augmented signal space $S \times \{1,0\}$. This means that the DM observes a binary signal in addition to $s'$. Note that this binary signal does not convey any additional information about the state given $s'$, i.e., $\Pi'$ is sufficient for this augmented experiment. 

We define conditional informativeness as follows.
 
\begin{defn}[\textbf{Conditional Informativeness}]
	\label{defn:n2}
	$\Pi' = (S^\prime, \{\pi^\prime_\theta\}_{\theta \in \Theta})$ is \emph{conditionally more informative} than $\Pi = (S, \{\pi_\theta\}_{\theta \in \Theta})$ with probability $\alpha \in (0,1]$ if there exists $\eta:S^\prime \rightarrow [0,1]$ such that $\left(S^\prime, {\pi^{\prime}_\theta }^\eta( \cdot | \{1\})\}_{\theta \in \Theta}\right)$ is more informative than $\Pi$ and ${\pi^{\prime}_\theta }^\eta(S^\prime \times \{1\}) = \alpha$ for every $\theta \in \Theta$. $\Pi'$ is \emph{$\alpha^*$-conditionally more informative} than $\Pi$ if $\alpha^* \in (0,1]$ is the supremum over $\alpha$ such that $\Pi'$ is conditionally more informative than $\Pi$ with probability $\alpha$.
	 
\end{defn}

In words, $\Pi^\prime$ is conditionally more informative than $\Pi$ if we can generate an event based on $s'$ with probability $\alpha$ independent of $\theta$, conditional on which $\Pi^\prime$ is more informative than $\Pi$ in the standard sense.\footnote{It is without loss of generality to augment $S'$ with two points.} Note that this event does not provide any information about the state. When the conditioning event $1$ occurs with probability 1 (i.e. $\alpha =1$), this definition reduces to the standard Blackwell informativeness.

If $\Pi^\prime$ is conditionally more informative than $\Pi$ with probability $\alpha$, then clearly it is conditionally more informative than $\Pi$ with any probability $\tilde{\alpha} < \alpha$, with different $\eta$.\footnote{Replace $\eta$ with $\eta \frac{\tilde{\alpha}}{\alpha}$.} In general, the probability of the conditioning event may differ across different $\eta$ functions. So, $\alpha^*$ is the supremum of the probability of such a conditioning event, given which $\Pi'$ becomes Blackwell more informative than $\Pi$.

It is not difficult to see that weighted garbling and conditional informativeness are very closely related. For example, in \Autoref{exmp:1}, $\Pi$ is a weighted garbling of $\Pi^\prime$ with weight $\gamma_{s_1^\prime} = \gamma_{s_2^\prime} =2$. At the same time, $\Pi^\prime$ is more informative than $\Pi$ conditional on $s_1^\prime$ or $s_2^\prime$ being observed, which occurs with probability $0.5$ independent of state. 

We can establish this equivalence for the general finite signal case as follows. Suppose that $\Pi$ is a weighted garbling of $\Pi^\prime$ with weight $\gamma$. If we divide both sides of \eqref{eq:nnn1} in \autoref{defn:WG} by $\overline{\gamma}$, then we obtain the following expression for all $s \in S$: 
$$\frac{1}{\overline{\gamma}} \pi_\theta (s) = \sum_{ s' \in S'} \phi (s|s')\frac{\gamma_{s'}}{\overline{\gamma}} \pi'_\theta (s'), \quad \forall \theta \in \Theta.$$
We can interpret $\frac{\gamma_{s'}}{\overline{\gamma}} \in [0,1]$ as the probability of the conditioning event given $s'$ and set $\eta_{s'} = \frac{\gamma_{s'}}{\bar{\gamma}}$. Note that the probability of this event is the same, equal to $\frac{1}{\overline{\gamma}}$, for every state as follows:
$$ 
\sum_{s^\prime \in S^\prime } \frac{\gamma_{s'}}{\overline{\gamma}} \pi^\prime_\theta (s^\prime) =\frac{1}{\overline{\gamma}}.
$$
If we divide both sides by $\frac{1}{\overline{\gamma}}$, then for all $s \in S$.
$$\pi_\theta (s) = \sum_{ s' \in S'} \phi (s|s')\frac{\gamma_{s'}/\overline{\gamma}}{1/\overline{\gamma}} \pi'_\theta (s'), \quad \forall \theta \in \Theta.$$
This implies that $\Pi$ becomes a garbling of $\Pi'$ given the conditioning event, hence $\Pi'$ is conditionally more informative than $\Pi$ with probability $\frac{1}{\overline{\gamma}}$.

We can also show the converse: if $\Pi'$ is conditionally more informative than $\Pi$ with probability $\alpha$, then $\Pi$ is a weighted garbling of $\Pi'$, where weight $\gamma_{s'}$ is given by the ratio of the probability of observing $1$ given $s'$ to $\alpha$. 

The following proposition generalizes the above observations to the case with a more general signal space.

\begin{prop}
	\label{prop:n1}
	Let $\Pi = (S,\{\pi_\theta\}_{\theta \in \Theta})$ and $\Pi' = (S', \{\pi^\prime_\theta\}_{\theta \in \Theta})$ be experiments. Then the following hold: 
	\begin{enumerate}
		\item If $\Pi$ is a weighted garbling of $\Pi'$ with weight $\gamma \in \Gamma_{S'}$, then $\Pi'$ is conditionally more informative than $\Pi$ with probability $\frac{1}{\overline{\gamma}}$ with $\eta: S' \rightarrow [0,1]$, where $\eta_{s^\prime} = \frac{\gamma_s}{\overline{\gamma}}$. 
		\item If $\Pi'$ is conditionally more informative than $\Pi$ with probability $\alpha \in (0, 1]$ with $\eta: S' \rightarrow [0,1]$, then $\Pi$ is a weighted garbling of $\Pi^\prime$ with weight $\gamma \in \Gamma_{S'}$, where $\gamma_{s^\prime} = \frac{\eta_{s^\prime}}{\alpha}$.	
	\end{enumerate}
	
\end{prop}

\begin{proof}
	See \Cref{sec:app}.
\end{proof}

This implies a tight relationship between the size of weighted garbling and the maximum conditioning probability for conditional informativeness.

\begin{cor}
\label{cor:o1}
$\Pi$ is a weighted garbling of $\Pi^\prime$ with size $\beta$ if and only if $\Pi$ is $\alpha^*$-conditionally more informative than $\Pi'$ where $\alpha^* = \frac{1}{\beta}$.
\end{cor}
	
\begin{proof}
It is clear from the first item of the above proposition that $\alpha^* \geq \frac{1}{\beta}$. On the other hand, if $\Pi'$ is conditionally more informative with probability $\alpha$ and $\eta$ such that $\bar{\eta} = \sup_{s'} \eta_{s'} < 1$, then we can increase the probability of the conditioning event proportionally by $\frac{1}{\bar{\eta}}$ with weight $\tilde \eta_{s'} = \frac{\eta_{s'}}{\bar{\eta}}$. Hence we can focus on the class of $\eta$ such that $\bar{\eta} =1$ to determine $\alpha^*$. Then, the weight $\gamma$ in the second item of the proposition satisfies $\overline{\gamma} = \frac{1}{\alpha}$ with such $\eta$. This implies $\beta \leq \frac{1}{\alpha^*}$.
\end{proof}

\subsection{Weighted Garbling as Payoff Guarantee}

Next, we show that weighted garbling can be interpreted in terms of some fractional payoff guarantee. What we require is that the marginal value of information of one experiment is at least a fraction of the marginal value of information of another experiment for any decision problem. We show that this order is equivalent to the weighted garbling order, with a one-to-one relationship between the required fraction and the size of WG-relationship.  Let $V^{\mathcal{A}} (\varnothing)$ be the optimal expected payoff given the prior without any additional information.

\begin{thm}
\label{thm:n1}
Let $\Pi$ and $\Pi'$ be experiments. Then, for any $\beta \in [1,\infty)$, $\Pi$ is a weighted garbling of $\Pi'$ with size $\beta$ if and only if 
$$\inf_{ \mathcal{A}}  \frac{V^{ \mathcal{A}} (\Pi') - V^{ \mathcal{A}} (\varnothing)}{V^{ \mathcal{A}} (\Pi) - V^{ \mathcal{A}} (\varnothing)} = \frac{1}{\beta },$$ 
with the conventions $\frac{0}{0}=1$ and $\frac{c}{0} = \infty$ for $c > 0$.
\end{thm}

\begin{proof}
	See \Autoref{sec:app}.
\end{proof}

$V^{\mathcal{A}}(\Pi) - V^{ \mathcal{A}} (\varnothing)$ is the \emph{marginal value of information} of experiment $\Pi$, i.e., the difference between the optimal expected utilities with and without $\Pi$. 
The result implies that $\Pi$ is a weighted garbling of $\Pi'$ with size $\beta$ if and only if $\frac{1}{\beta}$ is the tight lower bound for the ratio of the marginal value of information of $\Pi'$ over the marginal value of information of $\Pi$ over all decision problems, i.e., this ratio is at least $\frac{1}{\beta}$ and can be made arbitrarily close to $\frac{1}{\beta}$ by choosing an appropriate decision problem. When $\beta$ is $1$, this reduces to the standard case in \cite{Blackwell_1951,Blackwell_1953_AMS}.

Note that if $\Pi$ is a weighted garbling of $\Pi'$ with weight $\gamma$, then the infimum of the ratio is at least $\frac{1}{\overline{\gamma}}$, since $\overline{\gamma}$ is at least as large as $\beta$. Thus, even when the exact size is unknown, the result provides a uniform lower bound on the ratio. Conversely, if the infimum of the ratio is bounded below by some $\frac{1}{\beta'}$, then $\Pi$ must be a weighted garbling of $\Pi'$ with size at most $\beta'$. Another implication of this result is that if $\Pi$ is not a weighted garbling of $\Pi'$, then for any $\beta'>0$, there exists a decision problem for which the marginal value of information of $\Pi$ exceeds $\beta'$ times the marginal value of information of $\Pi'$ (equivalently, the infimum of the ratio becomes 0).

The result is driven by two observations together with the definition of the size of a weighted garbling relationship: (i) If $\Pi$ is a weighted garbling of $\Pi^\prime$ with weight $\gamma$, then $\Pi^\prime$ is conditionally more informative than $\Pi$ with probability $\frac{1}{\overline{\gamma}}$ by \autoref{prop:n1}. When the conditioning event occurs, the marginal value of information of $\Pi'$ is at least that of the marginal value of information of $\Pi$.  When it does not occur, the DM can still secure $V^{\mathcal{A}}(\varnothing)$ by playing an ex-ante optimal action. Thus, $V^{\mathcal{A}}(\Pi') - V^{ \mathcal{A}} (\varnothing) \geq \frac{1}{\overline{\gamma}}  (V^{\mathcal{A}}(\Pi)-V^{\mathcal{A}}(\varnothing)) \geq \frac{1}{\beta}  (V^{\mathcal{A}}(\Pi)-V^{\mathcal{A}}(\varnothing))$. Hence, the infimum of the ratio is at least $\frac{1}{ \beta}$. (ii) Conversely, suppose the infimum of the ratio equals $\frac{1}{\beta}$. Consider the auxiliary experiment in which the DM observes a signal from $\Pi$ with probability $\frac{1}{ \beta}$ and observes nothing with the remaining probability. Note that $\Pi$ is a weighted garbling of this auxiliary experiment with size $\beta$. Also note that, by Blackwell's theorem, this auxiliary experiment is a garbling of $\Pi'$ (hence a weighted garbling of $\Pi'$ with size $1$), since $\Pi'$ is Blackwell more informative than this auxiliary experiment. Then the transitivity result in \autoref{prop:n2} implies that $\Pi$ is a weighted garbling of $\Pi'$ with some weight $\gamma$ satisfying $\overline{\gamma} \leq \beta$. Hence, the size is at most $\beta$. Combining (i) and (ii) and using the definition of size establishes the equalities in both directions: if $\Pi$ is a weighted garbling of $\Pi'$ with size $\beta$, then the infimum equals $\frac{1}{\beta}$, and conversely, if the infimum equals $\frac{1}{ \beta}$, then the size is $\beta$.

\vspace{3mm}

\textbf{Revisit \autoref{exmp:1}} We apply this result to \autoref{exmp:1}. In light of \autoref{cor:o1}, instead of obtaining the size directly,  we obtain the maximum $\alpha$ for which $\Pi'$ is conditionally more informative than $\Pi$ with probability $\alpha$. We claim that it is $\frac{1}{2} + \frac{1}{2} \min \{\frac{q' - q}{q- \frac{1}{2}}, 1\}$. To see this, suppose that the signal $s_0'$ is artificially converted into $s_1'$ or $s_2'$ randomly with probability $\eta$. If $\eta$ is small, then this experiment is still Blackwell-more informative than $\Pi$ conditional on $\{s_1', s_2'\}$. For the largest such $\eta$ that preserves Blackwell order in this way, this garbled experiment of $\Pi'$ becomes an experiment that is $\Pi$ with probability $\alpha^*$ and a null experiment with probability $1- \alpha^*$. 

This maximum $\eta$ is given by 
$$\frac{\frac{1}{2} q' + \frac{1}{2}\eta \times \frac{1}{2}}{\frac{1}{2} + \frac{1}{2} \eta} = q,$$ 
where the left-hand side is the probability of observing the correct signal given the conditioning event in each state. Then, the maximum conditioning probability $\alpha^*$ is given by $\frac{1}{2} + \frac{1}{2}  \eta$. For instance, when $q =\frac{2}{3}$ and $q' = \frac{3}{4}$, we have $\eta=\frac{1}{2}$ and $\alpha^* = \frac{3}{4}$.\footnote{The corresponding weights are $\gamma_{s_1'} = \gamma_{s_2'} = \frac{1}{\frac{1}{2} + \frac{1}{2} \eta}= \frac{4}{3}$ and $\gamma_{s_0'} = \frac{\eta}{\frac{1}{2} + \frac{1}{2}\eta}= \frac{2}{3}$.}  \autoref{thm:n1} then says that 
$$\inf_{\mathcal{A}} \frac{V^{\mathcal{A}} (\Pi') - V^\mathcal{A} (\varnothing)}{ V^{\mathcal{A}}(\Pi) - V^\mathcal{A} (\varnothing)} = \frac{3}{4},$$
with the conventions $\frac{0}{0}=1$ and $\frac{c}{0}=\infty$ for $c> 0$.

%% file: section5.tex
Weighted garbling captures the idea that a more informative signal can be observed probabilistically. Hence, it is natural to expect that a more WG-informative experiment would be more useful in dynamic settings where a DM can conduct the same experiment repeatedly. In this section, we consider the dynamic setting where the state is changing over time and a DM can conduct an experiment many times before making a choice. In the following, we make several simplifying assumptions on the primitives: $A$ is compact, $u$ is continuous, and all experiments are finite experiments. We normalize $u$ so that $u(a, \theta) \in \left[0,1\right]$ for every $(a, \theta) \in A \times \Theta$ without loss of generality.

 We use a hidden Markov process to model the state dynamics. Time is discrete and denoted by $t=0,1,\dots, T$. Let $\rho: \Theta \rightarrow \Delta(\Theta)$ be the transition kernel that generates a sequence of states $\theta_t, t=0,1,2,\dots,T$ given the initial state $\theta_0$. We denote the probability of $\theta_{t+1} = \theta^\prime$ given $\theta_{t} = \theta$ by $\rho(\theta^\prime| \theta)$ for any $\theta^\prime, \theta \in \Theta$. We assume that this Markov process is irreducible and aperiodic. Let $R$ be the set of all irreducible and aperiodic $\rho$. The dynamic decision problem $\left(\mathcal{A}, \rho, T \right)$ consists of a decision problem $\mathcal{A} = \left(A, u, \mu_0 \right)$, where $\mu_0 \in \Delta(\Theta)$ is the initial distribution over states, a transition kernel $\rho$, and terminal period $T$. In each period, the DM conducts an experiment $\Pi = \left(S, \{ \pi_\theta \}_{ \theta \in \Theta } \right )$, then decides whether to take an irreversible action once and for all or to wait and proceed to the next period. If the DM chooses $a \in A$ in period $t$, then she receives $u(a, \theta_t)$ and exits. The DM must make a choice by period $T$. There is no discounting, so waiting is not costly. Note that the DM may wait for two reasons: to acquire additional information and to allow the state to transition to a more favorable state. 

Formally, the DM's policy $f$ is a mapping from $\bigcup_{t=0}^T H_t$ to $\Delta(A_t)$, where $H_{t} = S^{t+1}$ is the set of $t+1$ signals $s^t = (s_0,\dots,s_{t})$, $A_t =A \cup \{w\}$ for $t=0,\dots,T-1$ and $A_T  =A$, and $w$ denotes the option of waiting. Let $F$ be the set of all such policies. The DM's expected payoff from $f \in F$ is 
\[
V^{\mathcal{A}, \rho, T}(f; \Pi) = \mathbb{E}^{\rho, \mu_0,  \Pi}\left[ \sum_{a_0 \in A} u\left(a_0, \theta_0 \right) f(s_0)(a_0) \\
 + \sum_{t=1}^T \sum_{s^{t}} \left(\prod_{n=0}^{t-1} f(s_0,\dots,s_n)(w)\right) \sum_{a_t \in A} u\left(a_t, \theta_t \right) f(s^t)(a_t)\right].
\]
Let $V^{\mathcal{A}, \rho, T}(\Pi) = \max_{f \in F}V^{\mathcal{A}, \rho, T}(f ; \Pi)$ be the DM's optimal payoff for the dynamic decision problem $\left(\mathcal{A}, \rho, T \right)$ given experiment $\Pi$.

We would like to introduce a few definitions regarding experiments to state our results. We say that $\Pi$ satisfies the \emph{bounded belief property} if $\pi_\theta (s) > 0$ for every $s \in S$ and $\theta \in \Theta$. This means that the support of posterior beliefs never changes over time after any finite sequence of signals. We also say that $\Pi$ and $\Pi'$ is a \emph{regular} pair of finite experiments if, for any full-support prior, no posterior belief for one experiment is not on the relative boundary (vertice, edges, faces etc.) of the convex hull of the posterior beliefs for the other experiment.\footnote{For any set $C$ in $\mathbb{R}^n$, its relative interior $ri(C)$ is the set of point $x$ in the affine full of $C$ for which there exists $\epsilon >0$ such that the intersection of open ball $B_\epsilon(x)$ and the affine hull of $C$ is in $C$. Then the relative boundary of $C$ is given by $cl C/ri(C)$, where $cl C$ is the closure of $C$ (\cite{Rockafellar_1997}, P.44).} For example, for the case of two states, a pair of experiments is regular if the maximum and minimum likelihood ratios (such as $\max_{s}\frac{\pi(s|\theta_1)}{\pi(s|\theta_2)}$) are different for them. For our result, we focus on the comparison of regular pairs of experiments, which are generic cases.

Our goal is to show that $\Pi$ is a weighted garbling of $\Pi^\prime$ if and only if, for a given transition kernel $\rho \in R$, there exists $T'$ such that $\Pi'$ performs better than $\Pi$ for dynamic decision problem $\left(\mathcal{A}, \rho, T\right)$ for any $\mathcal{A}$ and any $T \geq T'$.

To see why $\Pi^\prime$ would work better than $\Pi$ with repeated experiments, consider the special two-state i.i.d. case where \autoref{exmp:1} is repeated independently over time, where the state is redrawn every period according to the $50-50$ prior. In this case, the DM's posterior belief of $\theta_1$ or $\theta_2$ can be at most $q$ with $\Pi$, whereas it can be as high as $q' > q$ with $\Pi'$, although the signal is often uninformative for $\Pi'$. When the DM can repeat this experiment many times, the DM can act on more extreme beliefs with almost probability $1$, hence obtain a higher expected payoff with $\Pi'$.

For a more general case with changing states, we need to examine the set of posterior beliefs that are reachable in the long run more carefully. For each $\rho \in R$ and each experiment, we can show that there exists a unique ``recurrent belief set,'' which captures the set of all (extreme) posterior beliefs that would arise after conducting an experiment repeatedly in the long run. We can show that a WG-more informative experiment is indeed associated with a larger recurrent belief set, hence performs better when $T$ is large enough.

However, we need to focus on a restricted set of initial beliefs to obtain our characterization. For a WG-more informative experiment to be more useful, we need to exclude the possibility of temporary short-run beliefs that could arise only in the beginning. This is because the DM may not have enough chance of observing an informative signal in the short run, hence may not be able to take advantage of such short-run beliefs that are useful for making a better decision. So we focus on the set of not too extreme initial beliefs (relative to $\rho$ and experiments) where the posterior beliefs always stay within the recurrent belief set.

To state this assumption formally, we provide a formal definition of \emph{recurrent belief set} for the Markov process $\rho \in R$ given $\Pi$ as follows. Let $\eta_{\rho, \Pi}(E)$ be the convex hull of all possible posterior beliefs that could arise in the next period after a state transition and an experiment of $\Pi$ when the post-experiment belief in the current period belongs to $E \subset \Delta(\Theta)$.\footnote{The formal definition is $\eta_{\rho, \Pi}(E) = co\left(\{\mu^\prime|\exists \mu \in E, \ \exists s \in S, \ \forall \theta^\prime \in \Theta, \ \mu^\prime (\theta^\prime)= \frac{\pi_{\theta^\prime}(s)\sum_{\theta}\rho(\theta^\prime|\theta)\mu(\theta)}{\sum_{\theta^\prime}\pi_{\theta^\prime}(s)\sum_{\theta}\rho(\theta^\prime|\theta)\mu(\theta)}\}\right)$.} For any $\Pi$ with the bounded belief property and any closed and convex set $E$, this set is a closed and convex set. Since $\eta_{\rho, \Pi}(\Delta(\Theta)) \subset \Delta(\Theta)$, $\eta^n_{\rho, \Pi}(\Delta(\Theta)), n=1,2,\dots,$ is a nested sequence of closed and convex sets, the limit $\lim_{n \rightarrow \infty} \eta^n_{\rho, \Pi}(\Delta(\Theta))$ is a nonempty closed (hence compact) convex set. We call this set \emph{recurrent belief set} and denote it by $\eta_{\rho, \Pi}^\infty$. 
For any irreducible and aperiodic $\rho$, every belief in this set assigns positive probability to every state.

Suppose that $\Pi'$ is WG-more informative than $\Pi$. Then we make the following regularity assumption on initial belief $\mu_0$: $\supp(Q^{\Pi'}_{\mu_0}) \subset \eta_{\rho, \Pi^\prime}^\infty$, which means that the posterior belief after conducting experiment $\Pi'$ (hence after conducting experiment $\Pi$ as well) in period $0$ is in $\eta_{\rho, \Pi^\prime}^\infty$ with probability $1$. We denote the set of decision problems with the initial belief that satisfies this condition by $\mathcal{F}(\rho, \Pi^\prime)$. As an example, consider the i.i.d. state process: the state is redrawn each period according to a full-support distribution $\mu \in \Delta (\Theta)$ (this Markov process is irreducible and aperiodic). We denote such a process by $\rho^\mu$. In this example, $\eta^{\infty}_{\rho, \Pi'}$ for $\Pi'$ becomes the convex hull of the support of $Q_{\mu}^{\Pi'}$. Then this regularity condition is satisfied if and only if $\mu_0$ is $\mu$ itself. 

Now we can state our main result in this section.

\begin{thm}
\label{thm:4}
Let $\Pi$ and $\Pi^\prime$ be a regular pair of finite experiments that satisfy the bounded belief property. Then the following conditions are equivalent.
\begin{enumerate}
	\item $\Pi$ is a weighted garbling of $\Pi^\prime$.
	\item For any $\rho \in R$, there exists $T^\prime \in \mathbb{N}$ such that $V^{\mathcal{A}, \rho, T}(\Pi^\prime) \geq V^{\mathcal{A}, \rho, T}(\Pi)$ for any $T \geq T'$ and any decision prolbem $\mathcal{A} \in \mathcal{F}(\rho, \Pi^\prime)$.
\end{enumerate}
\end{thm}

When we show $2 \rightarrow 1$, we assume that $\Pi$ is not a weighted garbling of $\Pi'$ and use the i.i.d. process $\rho^\mu$ to show that $2$ does not hold. So, we obtain the following result for the i.i.d. case as a corollary.

\begin{cor}
Let $\Pi$ and $\Pi^\prime$ be a regular pair of finite experiments that satisfy the bounded belief property. Suppose that the transition is given by i.i.d. $\rho^\mu$ for some full-support distribution $\mu \in \Delta (\Theta)$. Then the following conditions are equivalent. 
\begin{enumerate}
	\item $\Pi$ is a weighted garbling of $\Pi^\prime$.
	\item There exists $T^\prime \in \mathbb{N}$ such that $V^{\mathcal{A}, \rho^\mu, T}(\Pi^\prime) \geq V^{\mathcal{A}, \rho^\mu, T}(\Pi)$ for any $T \geq T'$ and any decision problem $\mathcal{A}$ with $\mu_0 = \mu$. 
\end{enumerate} 
\end{cor}

The reason why the DM prefers $\Pi^\prime$ to $\Pi$ is that the set of dynamically reachable posterior beliefs is larger for $\Pi^\prime$. Recall that $\Pi^\prime \succeq_{WG} \Pi$ if and only if the set of posterior beliefs for $\Pi$ is included in the convex hull of posterior beliefs for $\Pi^\prime$ for any prior belief $\mu$. For a regular pair of $\Pi$ and $\Pi'$, this inclusion is strict for any full-support prior belief in the sense that the former set is in the relative interior of the latter set. We can show that this generalizes to dynamically reachable beliefs. The set of all posterior beliefs that could arise in the dynamic setting for $\Pi$ is included in the relative interior of the recurrent belief set $\eta_{\rho, \Pi^\prime}^\infty$ for $\Pi'$. Furthermore, any extreme point in the recurrent belief set is approximately reachable with an arbitrarily high probability in the long run (in a way that is independent of decision problem $\mathcal{A}$) when the bounded belief property is satisfied. This strict gap between the possible posterior beliefs given $\Pi$ and the recurrent posterior beliefs given $\Pi'$ is the source of $\Pi'$'s advantage over $\Pi$ in the long run.

%% file: section6.tex
For the optimal stopping problem in the previous section, we assume that the state changes over time. Suppose instead that the state is drawn only once according to some prior distribution and is fully persistent. Since there is no discounting, the DM does not have an incentive to make a decision before reaching the last experiment. Thus, the problem essentially becomes static: experiment $\Pi'$ performs better than experiment $\Pi$ for this problem if and only if there exists $T \in \mathbb{N}$ such that $T$ repetition of $\Pi'$ is more informative than $T$ repetition of $\Pi$ for any decision problem in the standard Blackwell sense. This information order is the large sample order, which is recently studied by \cite{Azrieli_2014_ECMA} and \cite{MPST_2021_ECMA}.\footnote{Azrieli \cite{Azrieli_2014_ECMA} referred to this order as ``eventually Blackwell sufficient.'' Here we follow the terminology of \cite{MPST_2021_ECMA}.} This suggests a connection between our WG-order and the large sample order.

In this section, we compare our WG-order to the large sample order.  Our main observation is that, for the special but important case with two states, which is the focus of \cite{MPST_2021_ECMA}, the large sample order implies WG-order. However, we can show by an example that they do not imply each other with three states or more. We also discuss their relation to another related extension of the Blackwell order introduced by \cite{MS_2002_ECMA}.

Given experiment $\Pi = (S, \{ \pi_{ \theta} \}_{ \theta \in \Theta})$ and $\Pi' = (S', \{ \pi'_{ \theta} \}_{ \theta \in \Theta})$, define $\Pi \otimes \Pi'$ as the experiment $(S \times S', \{\pi_\theta \times \pi'_{\theta} \}_{ \theta \in \Theta} )$, where $\pi_\theta \times \pi'_{\theta}$ denotes the product measure. Also, let $\Pi^{ \otimes n} \equiv \Pi \otimes \cdots \otimes \Pi$ ($n$ times). We say that $\Pi'$ is better than $\Pi$ in the large sample order if there exists $N \in \mathbb{N}$ such that for any $n \geq N$, $V^{ \mathcal{A}} (  \Pi^{ \prime  \otimes n} ) \geq V^{ \mathcal{A}} (  \Pi^{  \otimes n} )$ for any decision problem $\mathcal{A}$. \cite{MPST_2021_ECMA} provides a characterization of this order in terms of a one-shot experiment when $|\Theta | = 2$. We denote this order by $\succeq_{LD}$. If $N$ can depend on decision problem $\mathcal{A}$ in the above definition, then we obtain an order introduced by \cite{MS_2002_ECMA}, which we denote by $\succeq_{MS}$. More precisely, $\Pi' \succeq_{MS} \Pi$ if for any decision problem $\mathcal{A}$, there exists $N_{ \mathcal{A}} \in \mathbb{N}$ such that for any $n \geq N_{ \mathcal{A}}$, $V^{ \mathcal{A}} (  \Pi^{ \prime  \otimes n} ) \geq V^{ \mathcal{A}} (  \Pi^{   \otimes n} )$.\footnote{As \cite{MS_2002_ECMA} noted, $\succeq_{MS}$ is generically complete, unlike $\succeq_{LD}$ or $\succeq_{WG}$.}

The following proposition establishes a few (non-)relationships between these information orders.

\begin{prop}
\label{prop:n3}
Given two experiments $\Pi$ and $\Pi'$, the following hold:
\begin{enumerate}
	\item $\Pi' \succeq_{LD} \Pi$ does not imply $\Pi' \succeq_{WG} \Pi$. For $\Theta$ with $|\Theta| =2$, $\Pi' \succeq_{LD} \Pi$ implies $\Pi' \succeq_{WG} \Pi$.
	\item $\Pi' \succeq_{WG} \Pi$ does not imply $\Pi' \succeq_{LD} \Pi$.
\end{enumerate}
\end{prop}

\begin{proof}
	See \Cref{sec:app}.
\end{proof}

Since $\succeq_{LD}$ implies $\succeq_{MS}$ by definition, the result also implies that $\Pi' \succeq_{MS} \Pi$ does not imply $\Pi' \succeq_{WG} \Pi$. In fact, this is the case even when $|\Theta | =2$ (unlike $\succeq_{LD}$). In addition, we can show that $\Pi' \succeq_{WG} \Pi$ does not imply $\Pi' \succeq_{MS} \Pi$ either. Hence, the above non-relationship holds for WG-order and MS-order as well.  

This result suggests that the two extensions $\succeq_{LD}$ and $\succeq_{MS}$ of the Blackwell order and the WG-order $\succeq_{WG}$ capture different notions of informativeness.

In the rest of this section, we present an example with three states where $\Pi' \not \succeq_{WG} \Pi$, but $\Pi'$ ``dominates $\Pi$ in large samples in the WG order,'' i.e., there exists $N \in \mathbb{N}$ such that for any $n \geq N$, $\Pi^{ \prime \otimes n} \succeq_{WG} \Pi^{ \otimes n}$. This example is used in an intermediate step of the proof to show that $\succeq_{LD}$ does not imply $\succeq_{WG}$ (i.e., the first part of the first item of the proposition). Once we have such an example, we can mix $\Pi$ with the null experiment to strengthens $\Pi^{ \prime \otimes n} \succeq_{WG} \Pi^{ \otimes n}$ to $\Pi^{ \prime \otimes n} \succeq_{BG} \tilde{\Pi}^{ \otimes n}$ for all $n \geq N$, where $\tilde{\Pi}$ denotes such a mixture, which implies that $\Pi'$ dominates $\tilde \Pi$ in the large sample order. However, $\tilde{\Pi}$ is still not a weighted garbling of $\Pi'$ (i.e. $\Pi' \not \succeq_{WG} \tilde \Pi$) by \autoref{thm:2} because $\tilde{\Pi}$ and $\Pi$ still share the same convex hull of the support of the posteriors.

We like to note that this example is of independent interest, since it follows from our belief-based characterization that the convex hull of posterior beliefs from an experiment can include the posterior beliefs from the other experiment only after the same experiment is repeated multiple times, but not with a one-shot experiment.

\begin{lem}
\label{prop:pp3}
There exist experiments $\Pi = (S,\{ \pi_\theta \}_{ \theta \in \Theta} )$ and $\Pi' = (S', \{ \pi' \}_{ \theta \in \Theta} )$ such that 
\begin{enumerate}
\item $\Pi' \not \succeq_{WG} \Pi$
\item $\Pi^{ \prime \otimes 2} \succeq_{WG} \Pi^{ \otimes 2} $  and 
\item $\Pi^{ \prime \otimes 3} \succeq_{WG} \Pi^{ \otimes 3} $
\end{enumerate}
\end{lem}

\begin{proof}

Let $\Theta = \{ \theta_1, \theta_2, \theta_3\}$. For $\epsilon \in \left[0,\frac{1}{2} \right]$, let $\Pi^\epsilon = (S,  \{ \pi^\epsilon_\theta \}_{ \theta \in \Theta } )$ and $\Pi'=(S', \{ \pi'_\theta\}_{ \theta \in \Theta}  )$ be experiments, where $S = \{ s_1, s_2\}, S' = \{ s_1', s_2', s_3'\}$ and $\{ \pi^\epsilon_\theta\}_{ \theta \in \Theta}$ and $\{ \pi'_\theta \}_{ \theta \in \Theta}$ are described in \autoref{fig:01}. Note that $s \in S$ from $\Pi$ cannot distinguish $\theta_2$ and $\theta_3$. In fact, when $\epsilon =0$, $\Pi^\epsilon$ can be obtained from $\Pi'$ by ``merging'' $s_2'$ and $s_3'$ into $s_2$, and is otherwise identical to $\Pi'$.

\begin{figure}[t!]
\center
\begin{tabular}{|c|c|c|c|}
\hline
$\pi^\epsilon_\theta$ & $\theta_1$                                           & $\theta_2$                                           & $\theta_3$               \\ \hline
$s_1$ & $\frac{1}{2} + \epsilon$   & $\frac{1}{2} \left(\frac{1}{2} + \epsilon \right)$   & $\frac{1}{2} \left(\frac{1}{2} + \epsilon \right)$    \\ \hline
$s_2$ & $\frac{1}{2} - \epsilon$ & $1-\frac{1}{2} \left(\frac{1}{2} + \epsilon \right)$ &  $1-\frac{1}{2} \left(\frac{1}{2} + \epsilon \right)$\\ \hline
\end{tabular}
\quad 
\begin{tabular}{|c|c|c|c|}
\hline
$\pi'_\theta$      & $\theta_1$    & $\theta_2$    & $\theta_3$    \\ \hline
$s_1'$ & $\frac{1}{2}$ & $\frac{1}{4}$ & $\frac{1}{4}$ \\ \hline
$s_2'$ & $\frac{1}{4}$ & $\frac{1}{2}$ & $\frac{1}{4}$ \\ \hline
$s_3'$ & $\frac{1}{4}$ & $\frac{1}{4}$ & $\frac{1}{2}$ \\ \hline
\end{tabular}
\caption{}
\label{fig:01}
\end{figure}

\begin{figure}[t!]
\centering
		\begin{tikzpicture}
		\begin{ternaryaxis}[grid=none, ticks=none, clip=false,  every axis plot/.append style={mark size=1pt}]

\node at (axis cs:1,0,0) [anchor=south] {$\theta_1$};
\node at (axis cs:0,1,0) [anchor=east] {$\theta_2$};
\node at (axis cs:0,0,1) [anchor=west] {$\theta_3$};

		\addplot3[blue, style=dashed] coordinates {(1/2,1/4,1/4) (1/4,1/2,1/4) (1/4,1/4,1/2) (1/2,1/4,1/4)};
		\addplot3[only marks, red, mark=square*] coordinates {(1/2,1/4,1/4) (49/198,149/396,149/396)};
		\addplot3[blue, mark=*, style=dashed] coordinates {(2/3,1/6,1/6) (1/6,2/3,1/6) (1/6,1/6,2/3) (2/3,1/6,1/6)};
		\addplot3[only marks, red, mark=square*] coordinates {(2/3,1/6,1/6) (2/11,9/22,9/22)};
		\end{ternaryaxis}

	\end{tikzpicture}
\caption{Posterior beliefs from one-shot and twice-repeated experiments. The blue points are the (extreme) posterior beliefs associated with $\Pi'$, whereas the red points are those associated with $\Pi^\epsilon$ with $\epsilon = \frac{1}{100}$. In particular, $q_\epsilon(s_2) = \left( \frac{49}{198}, \frac{149}{396}, \frac{149}{396} \right).$ 
}
\label{fig:o1}
\end{figure}

We first show that for any $\epsilon >0$, $\Pi^\epsilon$ is not a weighted garbling of $\Pi'$. Assume that the prior belief is $(\frac{1}{3}, \frac{1}{3}, \frac{1}{3})$. For each $s' \in S'$, let $q' (s') \in \Delta (\Theta)$ be the posterior belief after observing $s'$. A direct calculation yields $q'(s_1') = \left(\frac{1}{2}, \frac{1}{4}, \frac{1}{4} \right)$, $q'(s_2') = (\frac{1}{4}, \frac{1}{2}, \frac{1}{4} )$, $q'(s_3') = (\frac{1}{4}, \frac{1}{4}, \frac{1}{2} )$. Similarly for $\Pi$, $q(s_1) = (\frac{1}{2}, \frac{1}{4}, \frac{1}{4} )$ and 
$$q_\epsilon (s_2) = \left( \frac{1-2\epsilon}{4(1-\epsilon)} ,\frac{3-2\epsilon}{8(1-\epsilon)},\frac{3-2\epsilon}{8(1-\epsilon)} \right).$$
Note that $q_\epsilon (\theta_1 |s_2) < \frac{1}{4} = q_0 (\theta_1 |s_2) = q'(\theta_1 |s_2')= q' (\theta_1 |s_3')$, so $q_\epsilon(s_2) \notin co( \{ q' (s_1'), q'(s_2'), q'(s_3')\})$. Therefore, $\Pi^\epsilon$ is not a weighted garbling of $\Pi'$ (see \autoref{fig:o1}).

Now we show that $\Pi^{ \prime \otimes 2} \succeq_{WG} \Pi^{\epsilon \otimes 2}$ for sufficiently small $\epsilon$. Without loss of generality, assume that the prior belief is $( \frac{1}{3},  \frac{1}{3},  \frac{1}{3})$. For small $\epsilon$, the posterior belief after observing $(s_2, s_2)$ is close to $( \frac{2}{11}, \frac{9}{22}, \frac{9}{22})$, which is the posterior belief when $\epsilon =0$. On the other hand, for $\Pi'$, the equal-weight convex combination of the posterior beliefs induced by $(s_2', s_2')$ and $(s_3',s_3')$ is $( \frac{1}{6}, \frac{5}{12}, \frac{5}{12})$. Note that $( \frac{2}{11}, \frac{9}{22}, \frac{9}{22})$ lies on the line segment between $(\frac{2}{3},\frac{1}{6}, \frac{1}{6})$ (the posterior belief induced by $(s_1', s_1')$) and $(\frac{1}{6}, \frac{5}{12}, \frac{5}{12})$. This means that $( \frac{2}{11}, \frac{9}{22}, \frac{9}{22})$ is a convex combination of the posterior beliefs induced by $\Pi^{\prime \otimes 2}$; hence, for sufficiently small $\epsilon$, the posterior belief after observing $(s_2,s_2)$ is also such a convex combination. Note that $(s_1,s_1)$ and $(s_1',s_1')$ induce the same posterior belief. Thus, \autoref{thm:2} implies that $\Pi^{\epsilon \otimes 2}$ is a weighted garbling of $\Pi^{ \prime \otimes 2}$.

Next, we show that $\Pi^{ \prime \otimes 3} \succeq_{WG} \Pi^{\epsilon \otimes 3}$ for sufficiently small $\epsilon$. For $\Pi'$, the posterior belief after observing $(s_k', s_k', s_k')$ assigns probability $\frac{4}{5}$ to state $\theta_k$ and $\frac{1}{10}$ to each of the remaining states. For $\Pi^\epsilon$ with small $\epsilon$, the posterior belief after observing $(s_2, s_2, s_2)$ is close to $(\frac{4}{31}, \frac{27}{62}, \frac{27}{62})$ (the posterior belief when $\epsilon = 0$). Note that $\frac{1}{2}q' (s_2', s_2', s_2') + \frac{1}{2} q' (s_3', s_3', s_3') = (\frac{1}{10}, \frac{9}{20}, \frac{9}{20})$, and since $\frac{1}{10} < \frac{4}{31} < \frac{4}{5}$, $q_\epsilon (s_2, s_2, s_2)$ is a convex combination of $\{q' (s_k', s_k', s_k')\}_{k=1,2,3}$ for sufficiently small $\epsilon$. By \autoref{thm:2}, this implies that $\Pi^{\epsilon \otimes 3}$ is a weighted garbling of $\Pi^{ \prime \otimes 3}$. 
\end{proof}
The intuition behind the example is as follows: $\Pi'$ has signals $s_k', k=1,2,3$ suggesting the state being $\theta_k$. $\Pi^\epsilon$ is a variant of $\Pi'$ in which $s_2'$ and $s_3'$ are merged as $s_2$ with $\epsilon$ perturbation that makes it slightly more informative about whether the true state is $\theta_1$ or not than $\Pi'$. Hence, $\Pi^\epsilon$ is not a weighted garbling of $\Pi'$. For two samples, observing $(s_2, s_2)$ from $\Pi$ can be regarded as merging of four different two-sample observations from $\Pi'$: $(s_2', s_2'), (s_2', s_3'), (s_3', s_2')$ and $(s_3',s_3')$. We observe that $(s_2', s_2')$ and $(s_3', s_3')$ yield a smaller posterior belief for $\theta_1$ than the other two: $(s_2', s_2')$ and $(s_3', s_3')$ strongly suggest $\theta_2$ and $\theta_3$, respectively, and the posterior belief for $\theta_1$ is $\frac{1}{6}$. On the other hand, the posterior belief for $\theta_1$ given $(s_2', s_3')$ or $(s_3', s_2')$ is $\frac{1}{5}$. The two samples of $s_2$ from $\Pi$ yields the posterior belief for $\theta_1$ between $\frac{1}{6}$ and $\frac{1}{5}$. 
 
Note that such an example is not available when $|\Theta| = 2$, because if an experiment has a more extreme posterior for either state with one sample, it also has a more extreme posterior belief for the state with multiple samples.

%% file: section7.tex
In this study, we explore an information order over experiments based on a generalization of garbling, which we term weighted garbling. Our main results provide characterizations of this order in terms of two distinct classes of decision problems. For static Bayesian decision problems, one experiment is more informative than another in the weighted-garbling order if and only if a decision maker's value of information from the former is guaranteed to be a fixed fraction of the value of information from the latter for any decision problem. For a class of dynamic stopping time problems with a hidden Markov process, in which a patient decision maker can conduct an experiment as many times as desired at no cost before making a one-time decision, an experiment is more informative than another in the weighted-garbling order if and only if the decision maker achieves a weakly higher expected payoff for any problem with a regular prior belief in this class.

%% file: appendix.tex
\subsection{Proof of \autoref{prop:n2}}

\begin{proof}
	Suppose $\Pi = (S, \{\pi_\theta\}_{ \theta \in \Theta })$ is a weighted garbling of $\Pi' = (S', \{\pi'_\theta \}_{ \theta \in \Theta })$ with weight $\gamma_1 \in \Gamma_{S'}$ and $\Pi'$ is a weighted garbling of $\Pi'' =(S'',  \{\pi''_\theta \}_{ \theta \in \Theta} )$ with weight $\gamma_2 \in \Gamma_{S''}$. 
	Define $\phi: S'' \rightarrow \Delta(S)$ by 
	$$\phi (X |s'') : = \frac{\int_{S'} \gamma_{1,s'} \phi_1 (X|s') \phi_2 (ds' |s'')}{\int_{S'} \gamma_{1,s'} \phi_2 (ds'|s'')} $$ for each Borel set $X \subseteq S$ and $s'' \in S''$ such that $\int_{S'} \gamma_{1,s'} \phi_2 (ds'|s'') \neq 0$. For any other $s''$, let $\phi (\cdot |s'') \in \Delta (S)$ be any fixed measure. Then $\phi$ is a Markov kernel. 
	Also, define $\gamma: S'' \to \mathbb{R}_+$ by
	$$\gamma_{s''} : = \gamma_{2,s''} \int_{S'} \gamma_{1,s'} \phi_2 (ds' |s''), \quad \forall s'' \in S''.$$
	Observe that, for any Borel set $X \subseteq S$, 
	\begin{align*}
		&\int_{S''} \gamma_{s''} \phi (X|s'')\pi''_\theta (ds'')\\
		&\quad = \int_{S''} \gamma_{2,s''}\left( \int_{S'} \gamma_{1,s'} \phi_1 (X|s') \phi_2 (ds'|s'')  \right) \pi''_\theta (ds'')\\
		&\quad = \int_{S'} \int_{S''}  \gamma_{1,s'} \phi_1 (X|s') \gamma_{2,s''} \phi_2 (ds'|s'')  \pi''_\theta (ds'')\\
		&\quad =\int_{S'} \gamma_{1,s'} \phi_1 (X|s') \pi'_\theta(ds') \\
		&\quad = \pi_\theta(X).
	\end{align*}
	Therefore, $\Pi$ is a weighted garbling of $\Pi''$ with weight $\gamma \in \Gamma_{S''}$. By definition of $\gamma_{s''}$, we have $\gamma_{s''}\leq \gamma_{2,s''} \overline{\gamma}_1 \leq \overline{\gamma}_2 \overline{\gamma}_1$ for any $s'' \in S''$, since $\overline{\gamma}_1 = \sup_{s'} \gamma_{s'}$ and $\overline{\gamma}_2 = \sup_{s''} \gamma_{s''}$. Hence $\overline{\gamma} \leq \overline{\gamma}_2 \overline{\gamma}_1$.
	
\end{proof}

\subsection{Proof of \autoref{thm:1}}

\label{subsec:A1}

\begin{proof}

In the following proof, we use the following fact we observed previously. The proof is omitted. 

\begin{lem}\label{lem:1}
$\Pi =(S, \{\pi_\theta \}_{ \theta \in \Theta })$ is a weighted garbling of $\Pi'=(S', \{\pi'_\theta \}_{ \theta \in \Theta})$ with weight $\gamma \in \Gamma_{S'}$ if and only if $\gamma \Pi^\prime$ is an experiment (i.e. $\gamma \pi'_\theta \in \Delta(S')$ for every $\theta \in \Theta$) and $\Pi$ is a garbling of $\gamma \Pi^\prime$.
\end{lem}

We also assume, without loss of generality, that every experiment is a standard experiment, i.e., the signal space is $\Delta(\Theta)$. For a given prior, each signal for an experiment is equal to the posterior belief it generates. So, we denote a generic signal by $q \in \Delta(\Theta)$.

Take any full-support prior $\mu_0$. Suppose that $\Pi$ is a weighted garbling of $\Pi'$. 
By \autoref{lem:1}, $\Pi$ is a garbling of experiment $\gamma \Pi'$. Then, by Blackwell's theorem, $Q^{\gamma \Pi'}_{\mu_0} \in \Delta(\Delta(\Theta))$ is a mean-preserving spread of $Q^{\Pi}_{\mu_0} \in \Delta(\Delta(\Theta))$. Also note that 
$Q^{\gamma \Pi'}_{\mu_0}(X) = \sum_\theta \gamma \pi'_\theta(X) \mu_0(\theta) \leq \sum_\theta \overline{\gamma} \pi'_\theta(X) \mu_0(\theta) = \overline{\gamma} Q^{\Pi'}_{\mu_0}(X)$. Hence, $\sup_{X \in \mathcal{B}_{\Delta(\Theta)}} \frac{Q^{\gamma \Pi'}_{\mu_0}(X)}{Q^{\Pi'}_{\mu_0}(X)} \leq \overline{\gamma}$. Since $\overline{\gamma}$ is the essential supremum of $\gamma$, for any $\epsilon > 0$, there exists some positive probability event $X_\epsilon$ on which $\overline{\gamma}-\epsilon \leq \gamma_{q}$, hence $(\overline{\gamma}-\epsilon) Q^{\Pi'}_{\mu_0}(X_\epsilon) \leq \int_{X_\epsilon} \gamma_{q} Q^{\Pi'}_{\mu_0}(dq) = Q^{\gamma\Pi'}_{\mu_0}(X_\epsilon)$. This implies $\overline{\gamma}-\epsilon \leq \sup_{X \in \mathcal{B}_{\Delta(\Theta)}} \frac{Q^{\gamma \Pi'}_{\mu_0}(X)}{Q^{\Pi'}_{\mu_0}(X)}$. Since $\epsilon$ is arbitrary, we obtain equality $\sup_{X \in \mathcal{B}_{\Delta(\Theta)}} \frac{Q^{\gamma \Pi'}_{\mu_0}(X)}{Q^{\Pi'}_{\mu_0}(X)} = \overline{\gamma}$. Then setting $Q = Q^{\gamma \Pi'}_{\mu_0}$ proves the only if part.

To prove the second item, suppose that there exists a full-support prior $\mu_0$ and $Q \in \Delta(\Delta(\Theta))$ and $Q \ll Q^{\Pi'}_{\mu_0}$ such that $Q$ is a mean-preserving spread of $Q^{\Pi}_{\mu_0}$ and $\sup_{X \in \mathcal{B}_{\Delta(\Theta)}} \frac{Q(X)}{Q^{\Pi'}_{\mu_0}(X)} = \kappa$ for some $\kappa < \infty$. By absolute continuity, there exists Radon-Nikodym derivative $\gamma: S' \rightarrow \mathbb{R}$ to satisfy $Q(dq) = \gamma_q Q^{\Pi'}_{\mu_0}(dq)$. Clearly we can pick a nonnegative function for $\gamma$, hence $\gamma \in \Gamma_{S'}$. The essential supremum $\overline{\gamma}$ is at least as large as $\kappa$, since $Q(X) \leq \overline{\gamma} Q^{\Pi'}_{\mu_0}(X)$ for any Borel set $X$. Now we assume that $\overline{\gamma} > \kappa$ then derive a contradiction. Take $\epsilon> 0 $ such that $\overline{\gamma} - \epsilon > \kappa$. Since $\overline{\gamma}$ is the essential supremum, there exist a positive probability event $X_\epsilon$ for which $Q(X_\epsilon)  = \int_{X_\epsilon} \gamma_q Q^{\Pi'}_{\mu_0}(dq) \geq (\overline{\gamma}- \epsilon) Q^{\Pi'}_{\mu_0}(X_\epsilon)  > \kappa Q^{\Pi'}_{\mu_0}(X_\epsilon)$ holds. This is a contradiction, hence $\overline{\gamma} = \kappa$ must hold.

To show that $\Pi$ is a weighted garbling of $\Pi'$ with this $\gamma$, we first show that $\gamma \pi^\prime_\theta$ is a probability measure on $\Delta(\Theta)$ for each $\theta$, hence $\gamma \Pi^\prime =(\Delta(\Theta), \{\gamma  \pi^\prime_\theta\}_{ \theta \in \Theta})$ is indeed an experiment. For any $\theta \in \Theta$, $\mu_0(\theta) \pi^\prime_\theta$ is absolutely continuous with respect to $Q_{\mu_0}^{\Pi^\prime}$ where the Radon-Nikodym derivative is a posterior belief $q(\theta)$ on $\theta$.\footnote{Note that, for given $\theta$, $\mu_0(\theta) \pi^\prime_\theta(X) = \int_X q(\theta) Q_{\mu_0}^{\Pi^\prime} (dq)$ for any Borel Set $X \subseteq \Delta(\Theta)$.} Hence, for each $\theta$, 
\begin{align*}
	\int_{\Delta(\Theta)} \gamma_q \pi'_{\theta}(dq)
	&= \frac{1}{\mu_0(\theta)} \int_{\Delta(\Theta)}  \gamma_q \mu_0(\theta) \pi'_{\theta}(dq) \\ 
	&= \frac{1}{\mu_0(\theta)} \int_{\Delta(\Theta)} \gamma_q q(\theta) Q_{\mu_0}^{\Pi^\prime}(dq) \\
	&= 	\frac{1}{\mu_0(\theta)} \int_{\Delta(\Theta)} q(\theta) Q(dq).
\end{align*}
Since $Q$ is a mean preserving spread of $Q_{\mu_0}^{ \Pi}$,
$\int_{\Delta(\Theta)} q(\theta) Q(dq) = \int_{\Delta(\Theta)} q(\theta)Q_{\mu_0}^{ \Pi}(dq) = \mu_0(\theta)$ holds. Hence $\int_{\Delta(\Theta)} \gamma_q \pi'_{\theta}(dq) =1$ for every $\theta \in \Theta$.
Next we show that the posterior belief distribution generated by $\gamma \Pi^\prime$ is exactly $Q$. Remember that the posterior belief given signal $q$ is $q$ for $\Pi^\prime$ by definition. It is easy to see that the posterior belief given signal $q$ is $q$ for $\gamma \Pi^\prime$ as well. Hence, for any Borel set $X \subseteq \Delta(\Theta)$,
\begin{align*}
	\int_{X} Q^{\gamma \Pi^\prime}_{\mu_0} (dq) 
	&= \int_{X} \sum_\theta  \mu_0(\theta) \gamma_q \pi'_\theta(dq)\\
	&= \int_{X} \gamma_q Q^{\Pi^\prime}_{\mu_0}(dq)\\
	&=Q(X).
\end{align*}

Then it follows from Blackwell's theorem that experiment $\gamma \Pi^\prime$ is Blackwell more informative than $\Pi$, because $Q_{\mu_0}^{\gamma \Pi'} = Q$ is a mean preserving spread of $Q^{\Pi}_{\mu_0}$. By \autoref{lem:1}, $\Pi$ is a weighted garbling of $\Pi^\prime$ with weight $\gamma$. This proves the if part.
\end{proof}

\subsection{Proof of \autoref{prop:n1}}

\begin{proof}
Suppose $\Pi$ is a weighted garbling of $\Pi'$ with some weight $\gamma \in \Gamma_{S'}$. So the following holds for some Markov kernel $\phi: S' \rightarrow \Delta(S)$.
$$ \pi_\theta(X) = \int_X \gamma_{s'} \phi(X|s') \pi'_\theta(ds'),\quad  \forall X \in \mathcal{B}_{\Delta(S)}, \forall \theta \in \Theta$$

This condition is equivalent to the existence of $\alpha \in (0,1]$ and some function $\eta:S' \rightarrow [0,1]$ that satisfies,

\begin{equation}\label{eq:ci}
\alpha \pi_\theta(X) = \int_X \eta_{s'} \phi(X|s') \pi'_\theta(ds'), \quad \forall X \in \mathcal{B}_{\Delta(S)},\quad \forall \theta \in \Theta
\end{equation}
for the same $\phi$ via $\alpha = \frac{1}{\overline{\gamma}}$ and $\eta_{s'} = \frac{\gamma_{s'}}{\overline{\gamma}}$. 

For this $\eta$, it is easy to check that ${ \pi^{\prime}}^ \eta_\theta(S \times \{1\}) = \int_S \eta_{s'} \pi'_\theta(ds') = \alpha$ for every $\theta \in \Theta$. Also $\pi_\theta(X) = \int_X \phi(X|s') {\pi^{\prime}}^\eta_\theta(ds'|\{1\})$ holds for every Borel set $X \subseteq S$ for every $\theta \in \Theta$, where ${\pi^{\prime}}^\eta_\theta(ds'|\{1\}) = \frac{\eta_{s'}}{\alpha}\pi'_\theta(ds')$. This means that $\Pi$ is a garbling of $(S', \{{\pi^{\prime}}^\eta_\theta(\cdot| \{1\})\}_{\theta \in \Theta})$. Hence \eqref{eq:ci} is equivalent to $\Pi'$ being conditionally more informative than $\Pi$ with probability $\alpha$.

Then the second item of the proposition follows from \eqref{eq:ci} immediately by setting $\gamma_{s'} = \frac{\eta_{s'}}{\alpha}$.
\end{proof}

\subsection{Proof of \autoref{thm:n1}}

We prove the theorem by first establishing the following lemma.

\begin{lem} \label{lem:102}Let $\Pi$ and $\Pi'$ be experiments. Then the following hold.
\begin{enumerate}
\item Suppose $\Pi$ is a weighted garbling of $\Pi'$ with weight $\gamma \in \Gamma_{S'}$. Then $\inf_{ \mathcal{A}} \frac{V^{ \mathcal{A}} (\Pi') - V^{ \mathcal{A}} (\varnothing)}{V^{ \mathcal{A}} (\Pi) - V^{ \mathcal{A}} (\varnothing)} \geq \frac{1}{ \overline{\gamma}},$ with the conventions $\frac{0}{0}=1$ and $\frac{c}{0} =\infty$ for $c>0$.
\item Suppose $\inf_{ \mathcal{A}} \frac{V^{ \mathcal{A}} (\Pi') - V^{ \mathcal{A}} (\varnothing)}{V^{ \mathcal{A}} (\Pi) - V^{ \mathcal{A}} (\varnothing)} \geq \kappa$ for some $\kappa \in (0,1]$, with the conventions $\frac{0}{0}=1$ and $\frac{c}{0} = \infty$ for $c>0$. Then $\Pi$ is a weighted garbling of $\Pi'$ with some $\gamma \in \Gamma_{S'}$ such that $\overline{\gamma} \in \left[1,\frac{1}{ \kappa} \right]$.
\end{enumerate}
\end{lem}

\begin{proof}

\textbf{Proof of Item 1:} Let $\tilde{\Pi}^{ \prime \gamma} = (S', \{ \tilde{\pi}^{ \prime \gamma}_\theta \}_{ \theta \in \Theta} )$ be an experiment, where $\tilde{\pi}^{ \prime \gamma}_\theta  (X)  := \int_{X} \frac{1- \frac{\gamma_{s'}}{\overline{\gamma}} }{1- \frac{1}{ \overline{\gamma} }}\pi'_\theta (ds') $ for any Borel set $X \subseteq  S'$ and $\theta \in \Theta$.

Let $\sigma: S \to \Delta (A)$ be an arbitrary strategy under $\Pi$, and let $\sigma' : S' \to \Delta (A)$ be an arbitrary strategy under $\tilde{\Pi}^{\prime  \gamma}$. Define $\bar{\sigma}: S' \to \Delta (A) $ as follows: for each $s' \in S'$ and each Borel set $X \subseteq A$,
$$\bar{\sigma} (X|s')  := \frac{\gamma_{s'}}{\overline{\gamma}} \int_{ S}  \sigma ( X|s)\phi (ds|s') + \left( 1- \frac{\gamma_{ s'}}{\overline{\gamma}}\right) \sigma' (X|s').$$ 
Observe that
\begin{align*}
U^{ \mathcal{A}} (\bar{\sigma}; \Pi')&= \sum_{\theta \in \Theta} \int_{ S'} u (\bar{\sigma} (s') , \theta) \pi'_\theta (ds') \mu_0 (\theta) \\
& = \frac{1}{\overline{\gamma}} \sum_{ \theta \in \Theta} \int_{ S} u(\sigma (s), \theta)\left ( \int_{S'} \gamma_{s'} \phi (ds|s') \pi'_\theta (ds') \right) \mu_0 (\theta) \\
&\qquad + \left( 1- \frac{1}{ \overline{\gamma}} \right)\sum_{ \theta \in \Theta}  \int_{S'} \left ( \frac{1- \frac{\gamma_{s'} }{ \overline{\gamma} } }{1- \frac{1}{ \overline{\gamma}}} \right) u (\sigma' (s'), \theta) \pi'_\theta  (ds' ) \mu_0 (\theta) \\
& =\frac{1}{\overline{\gamma}}\sum_{ \theta \in \Theta}  \int_{ S}  u(\sigma (s), \theta) \pi_\theta (ds)\mu_0 (\theta) + \left( 1- \frac{1}{ \overline{\gamma}} \right) \sum_{ \theta \in \Theta} \int_{ S'} \left ( \frac{1- \frac{\gamma_{s'} }{ \overline{\gamma} } }{1- \frac{1}{ \overline{\gamma}}} \right) u (\sigma' (s'), \theta) \pi'_\theta (ds') \mu_0 (\theta) \\
&  = \frac{1}{\overline{\gamma}}U^{\mathcal{A}} (\sigma;\Pi) + \left( 1- \frac{1}{ \overline{\gamma}} \right)  U^{\mathcal{A}} (\sigma'; \tilde{\Pi}^{ \prime \gamma}).
\end{align*}
Since $\sigma$ and $\sigma'$ were arbitrary, we have 
\begin{align*}
V^{ \mathcal{A}} (\Pi') &\geq \frac{1}{ \overline{\gamma}} V^{ \mathcal{A}} (\Pi) + \left( 1- \frac{1}{ \overline{\gamma}} \right) V^{ \mathcal{A}} (\tilde{\Pi}^{ \prime  \gamma}) \\
& \geq \frac{1}{ \overline{\gamma}} V^{ \mathcal{A}} (\Pi) + \left( 1- \frac{1}{ \overline{\gamma}} \right) V^{ \mathcal{A}} (\varnothing),
\end{align*}
where the second inequality follows because $V^{ \mathcal{A}} (\tilde{\Pi}^{ \prime \gamma}) \geq V^{ \mathcal{A}} (\varnothing)$.

\vspace{0.5cm}

\textbf{Proof of Item 2:} If $\Pi$ is totally uninformative, then $\Pi$ is clearly a garbling of $\Pi'$ (i.e., $\overline{\gamma} = 1$).

Suppose that $\Pi$ is informative. Let $\tilde{\Pi} := (S \cup \{ n\}, \{\tilde{\pi}_\theta \}_{\theta \in \Theta})$ be an experiment, where $\tilde{\pi}_\theta  (X) := \kappa \pi_\theta (X)$ for each Borel set $X \subseteq S$ and $\tilde{\pi}_\theta ( \{n \}) := 1- \kappa$ for each $\theta \in \Theta$. That is, $\tilde{\Pi}$ is the experiment in which the DM observes a signal from $\Pi$ with probability $\kappa$ and observes ``nothing'' with the remaining probability. Then $\Pi$ is a weighted garbling of $\tilde{\Pi}$ with $\gamma_{\tilde{s}} = \frac{1}{\kappa}$ for $\tilde{s} \in S$ and $\gamma_{n} =0$ (so $\overline{\gamma} = \frac{1}{\kappa}$).

Note that given a decision problem $\mathcal{A}$, 
$$V^{ \mathcal{A}} (\tilde{\Pi}) = \kappa V^{ \mathcal{A}} (\Pi) + \left( 1- \kappa \right) V^{ \mathcal{A}} (\varnothing).$$
By assumption, $V^{ \mathcal{A}} (\Pi') \geq V^{ \mathcal{A}} (\tilde{\Pi})$ for each decision problem $\mathcal{A}$, so $\Pi'$ is Blackwell more informative than $\tilde{\Pi}$. By Blackwell's result, $\tilde{\Pi}$ is a garbling of $\Pi'$ (equivalently, $\tilde{\Pi}$ is a weighted garbling of $\Pi'$ with $\gamma_{s'} =1$ for all $s' \in S'$.). Then, by \autoref{prop:n2}, $\Pi$ is a weighted garbling of $\Pi'$ with some $\gamma \in \Gamma_{S'}$ with $\overline{\gamma}  \in  \left [1,\frac{1}{\kappa}\right ]$.
\end{proof}

\begin{proof}[Proof of \autoref{thm:n1}]

\textbf{``Only if'' part:} Suppose that $\Pi$ is a weighted garbling of $\Pi'$ with size $\beta \in [1,\infty)$. Then for any $\epsilon >0$, $\Pi$ is a weighted garbling of $\Pi'$ with some $\gamma \in \Gamma_{S'}$ with $\overline{\gamma} \in [\beta, \beta+ \epsilon)$. Hence, by Item 1 of \autoref{lem:102}, $\inf_{ \mathcal{A}} \frac{V^{ \mathcal{A}} (\Pi') - V^{ \mathcal{A}} (\varnothing)}{V^{ \mathcal{A}} (\Pi) - V^{ \mathcal{A}} (\varnothing)} \geq \frac{1}{ \beta}$. To show that the infimum cannot be strictly larger than $\frac{1}{\beta}$, suppose instead that it equals $\frac{1}{\beta'}$ for some $\beta'<\beta$. Then Item 2 of \autoref{lem:102} implies that $\Pi$ is a weighted garbling of $\Pi'$ with some $\gamma$ with $\overline{\gamma} \leq \beta'$, contradicting the definitioin of $\beta$ as the size.

\textbf{``If'' part:} Suppose that $\Pi$ and $\Pi'$ satisfy the condition in the ``if'' part. Then Item 2 of \autoref{lem:102} implies that $\Pi$ is a weighted garbling of $\Pi'$ with size at most $\beta$. To show that the size cannot be strictly less than $\beta$, suppose otherwise. Then, by Item 1 of \autoref{lem:102}, the infimum of the ratio is strictly larger than $\frac{1}{\beta}$, a contradiction.
\end{proof}

\subsection{Proof of \autoref{thm:4}}

We need some special notations for this dynamic environment. Given any belief $\mu \in \Delta(\Theta)$ in the current period (after experiment $\Pi$), let $r_{\rho,\Pi}(s|\mu) \in \Delta(\Theta)$ be the posterior belief in the next period after a state transition according to $\rho$ and signal $s$ being observed with experiment $\Pi$. The formal definition is 
$$
r_{\rho,\Pi}(s| \mu)(\theta^\prime) = \frac{\sum_{\theta \in \Theta}\pi_{ \theta^\prime} (s)\rho(\theta^\prime| \theta)\mu(\theta)}{\sum_{\theta^\prime \in \Theta} \sum_{\theta \in \Theta}\pi_{\theta^\prime}(s)\rho(\theta^\prime| \theta)\mu(\theta)}, \quad \forall \theta' \in \Theta.
$$
Similarly, we define the posterior belief $r_{\rho,\Pi}(s^n| \mu)$ recursively by $r_{\rho,\Pi}(s^n| \mu) = r_{\rho,\Pi}\left(s_n| r_{\rho,\Pi}(s^{n-1}| \mu)\right)$ for any $s^n = (s^{n-1},s_n)$. Note that $r_{\rho,\Pi}\left(s^{n}|\mu\right)$ is continuous in $\mu$ given any $s^n \in S^n$ by the bounded belief property.

We need a few lemmas for the proof of the main theorem. In the following, we always assume that $\rho$ is in $R$ and every experiment satisfies the bounded belief property.

\begin{lem}\label{lem:D1}
$\eta_{\rho, \Pi}^\infty$ satisfies the following properties.
\begin{enumerate}
\item For any $\mu \in \eta_{\rho, \Pi}^\infty$ and $s \in S$, $r_{\rho, \Pi}(s | \mu) \in \eta_{\rho, \Pi}^\infty$.
\item For any extreme point $\mu$ of $\eta_{\rho, \Pi}^\infty$, there exists an extreme point $\mu^\prime$ of $\eta_{\rho, \Pi}^\infty$ and $s \in S$ such that $\mu = r_{\rho, \Pi}(s| \mu^\prime)$. 
\end{enumerate}
\end{lem}

\begin{proof}

Since $\mu \in \eta_{\rho, \Pi}^n(\Delta(\Theta)) \bigcap \Delta(\Theta)$ for any $n=1,2,\dots$, $r_{\rho, \Pi}(s|\mu) \in \eta_{\rho, \Pi}^{n+1}(\Delta(\Theta))$ for any $n=0,1,\dots$. Hence, $r_{\rho, \Pi}( s|\mu) \in \bigcap_n \eta_{\rho, \Pi}^n(\Delta(\Theta)) = \eta_{\rho, \Pi}^\infty$. This proves the first item.

Take any extreme point $\mu$ of $\eta_{\rho, \Pi}^\infty$. For any $n \in \mathbb{N}$, $\mu \in \eta_{\rho, \Pi}^{n+1}(\Delta(\Theta))$; hence, there exist $\mu_n^k \in \eta_{\rho, \Pi}^n(\Delta(\Theta))$ and $s_n^k \in S , k=1,\dots,K$ such that $\mu \in co\left(\{r_{\rho, \Pi}(s_n^k|\mu_n^k),k=1,\dots,K \}\right)$. $K$ can be at most $\left|\Theta\right|$ by Carath\'{e}odory's theorem and is assumed to be the same for every $n$ (we can duplicate the same $(\mu^k_n, s_n^k)$ if needed). Take a subsequence of such vectors $\left((s_n^1, \mu_n^1),\dots,(s_n^K, \mu_n^K)\right), n=1,\dots$ in such a way that $(s_{n(t)}^k,\mu_{n(t)}^k) \rightarrow (s^k, \mu^k) \in S \times \Delta(\Theta)$ and $\sum_{k=1}^K \alpha_{n(t)}^k r_{\rho, \Pi}(s_{n(t)}^k|\mu_{n(t)}^k) \rightarrow \sum_{k=1}^K \alpha^k r_{\rho, \Pi}(s^k|\mu^k) = \mu$ using the continuity of $r_{\rho, \Pi}$ (the subsequence is taken so that the weights $(\alpha_{n(t)}^1,\dots,\alpha_{n(t)}^K)$ converge to some $(\alpha^1,\dots,\alpha^K)$ as well).

Clearly, $\mu^k \in \eta_{\rho, \Pi}^\infty$ for each $k$; hence, $r_{\rho, \Pi}(s^k|\mu^k)$ is in $\eta_{\rho, \Pi}^\infty$ by the first item. Since $\mu$ is an extreme point, it must be that $\mu = r_{\rho, \Pi}(s^k|\mu^k)$ for any $k =1,\dots,K$. If $\mu^k$ is an extreme point of $\eta_{\rho, \Pi}^\infty$ for some $k$, then we are done. So suppose not. Take any $\mu^k$. Then we can find a set $\{ \mu^{\prime m} \}_{m=1}^M$ of extreme points of $\eta_{\rho, \Pi}^\infty$ and $\{ \alpha^m\}_{m=1}^M \subset \mathbb{R}_+$ such that $\mu^{k} = \sum_{m=1}^M \alpha^m \mu^{\prime m}$, since $\eta_{\rho, \Pi}^\infty$ is a compact and convex set.

Note that $\mu = r_{\rho, \Pi}(s^\prime|\mu^k)$ is a convex combination of $r_{\rho, \Pi}(s^\prime|\mu^{\prime m})$ for any $s^\prime$, because the following holds for any $\hat \theta \in \Theta$:
\begin{align*}
&r_{\rho, \Pi}(s^\prime|\mu^k) (\hat \theta) \\
&\quad = \frac{\pi_{\hat \theta}(s^\prime)\sum_{\theta \in \Theta}\rho(\hat \theta|\theta)\left(\sum_{m=1}^M \alpha^m \mu^{\prime m}(\theta)\right)}{\sum_{\theta^\prime \in \Theta} \pi_{\theta^\prime}(s^\prime)\sum_{\theta \in \Theta }\rho(\theta^\prime|\theta)\left(\sum_{m=1}^M \alpha^m \mu^{\prime m}(\theta)\right)} \\
&\quad = \sum_{m=1}^M \frac{\pi_{\hat \theta}(s^\prime)\sum_{\theta \in \Theta} \rho(\hat \theta|\theta) \mu^{\prime m}(\theta)}{\sum_{\theta^\prime \in \Theta}\pi_{\theta^\prime}(s^\prime) \sum_{\theta \in \Theta}\rho(\theta^\prime|\theta)\mu^{\prime m}(\theta)}\frac{ \sum_{\theta^\prime \in \Theta } \pi_{\theta^\prime}(s^\prime) \sum_{\theta \in \Theta} \rho(\theta^\prime|\theta) \alpha^m \mu^{\prime m}(\theta)}{\sum_{\theta^\prime \in \Theta} \pi_{\theta^\prime}(s^\prime)\sum_{\theta \in \Theta}\rho(\theta^\prime|\theta)\left(\sum_{m=1}^M \alpha^m \mu^{\prime m}(\theta)\right)} \\
&\quad = \sum_{m=1}^M r_{\rho, \Pi}(s^\prime|\mu^{\prime m})(\hat \theta) \frac{\sum_{\theta^\prime \in \Theta} \pi_{\theta^\prime}(s^\prime)\sum_{\theta \in \Theta} \rho(\theta^\prime|\theta)\alpha^m \mu^{\prime m}(\theta)}{\sum_{\theta^\prime \in \Theta} \pi_{\theta^\prime}(s^\prime)\sum_{\theta \in \Theta} \rho(\theta^\prime|\theta)\left(\sum_{m=1}^M \alpha^m \mu^{\prime m}(\theta)\right)}.
\end{align*}
Set $s^\prime = s^k$. Since $\mu = r_{\rho, \Pi}(s^k|\mu^k)$ is an extreme point of $\eta_{\rho, \Pi}^\infty$ and $r_{\rho, \Pi}(s^k|\mu^{\prime m}) \in \eta_{\rho, \Pi}^\infty$ by the first item, it must be that $\mu = r_{\rho, \Pi}(s^k|\mu^{\prime m})$ for every $m=1,\dots,M$. Thus we found a pair of $s^k$ and an extreme point in $\eta_{\rho, \Pi}^\infty$ that proves the second item in the lemma.
\end{proof}

Note that, as a corollary of this lemma, we obtain $\eta_{\rho, \Pi}\left(\eta^\infty_{\rho, \Pi}\right) = \eta^\infty_{\rho, \Pi}$, i.e. $\eta^\infty_{\rho, \Pi}$ is a fixed point of $\eta_{\rho, \Pi}$.

\begin{lem}\label{lem:D2}
For any $\epsilon> 0$, there exists $n \in \mathbb{N}$ such that, for any extreme point $\mu$ of $\eta_{\rho, \Pi}^\infty$, there exists a finite sequence of signals $s^n = (s_1,\dots,s_n)$ that satisfies $\left\|r_{\rho, \Pi}(s^n|\mu^\prime) -\mu \right\| < \epsilon$ for any $\mu^\prime \in \Delta(\Theta)$.
\end{lem}

\begin{proof}
For each $s \in S$, let $R_{\rho,\Pi}(s)$ be the $\left|\Theta\right| \times \left|\Theta\right|$ matrix that has $\pi_{\theta^\prime}(s) \rho(\theta^\prime| \theta)$ as its $(\theta, \theta^\prime)$-entry. Let $I_{\rho,\Pi}(s)$ be the $\left|\Theta\right| \times \left|\Theta\right|$ diagonal matrix where the diagonal element for the $\theta$-row is given by $\frac{1}{\Pr(s|\theta)}$, where $\Pr(s|\theta) = \sum_{\theta^\prime}\pi_{\theta^\prime}(s)\rho(\theta^\prime|\theta)$ is the probability of observing $s$ after a transition when starting from $\theta$. Then $I_{\rho,\Pi}(s) R_{\rho,\Pi}(s)$ is a $\left|\Theta\right| \times \left|\Theta\right|$ matrix where its $\theta$-row corresponds to the posterior belief given $s$ when starting from $\theta$. More generally, \text{redcolor}{let $\Pr(s^n|\theta)$ be the probability to observe a sequence of signals $s^n=(s_1,\dots,s_n)$ after $n$-times state transitions when the initial state is $\theta$ (this corresponds to the probability of observing $(s_1,\dots,s_n)$ from period $1$ to period $n$),} and let $I_{\rho,\Pi}(s^n)$ be the $\left|\Theta\right| \times \left|\Theta\right|$ diagonal matrix where the diagonal element for the $\theta$-row is given by $\frac{1}{\Pr(s^n|\theta)}$. Then the $\theta$-row of $I_{\rho,\Pi}(s^n) R_{\rho,\Pi}(s_1) \times \cdots  \times R_{\rho,\Pi}(s_n)$ corresponds to $r_{\rho, \Pi}(s^{n}|\mu_\theta)$: the posterior belief given $s^n$ when starting from $\theta$, where $\mu_\theta$ is the prior belief that puts probability 1 on $\theta$.

Since all the entries of $R_{\rho,\Pi}(s_i), i=1,\dots,n,$ are strictly positive by the bounded belief assumption, we can apply a technical lemma in \cite{PS_2012_RES} to obtain the following property:
For any $\epsilon>0$, there exists $N$ such that for every $n \geq N$ and every $s^{n} \in S^{n}$, every pair of rows of $I_{\rho,\Pi}(s^{n}) R_{\rho,\Pi}(s_1) \times \cdots  \times R_{\rho,\Pi}(s_n)$ (which are posterior beliefs given different initial states) is at most $\epsilon$ away from each other. That is, for any $\epsilon > 0$, we can choose $n$ large enough so that $\left\|r_{\rho,\Pi}(s^{n}|\mu_\theta) - r_{\rho,\Pi}(s^{n}|\mu_{\theta^\prime}) \right\|  < \epsilon$ for any $s^{n} \in S^{n}$ and any pair $(\theta, \theta^\prime)$. This implies that there exists $n$ for which $r_{\rho,\Pi}(s^{n}|\mu)$ and $r_{\rho,\Pi}(s^{n}|\mu^\prime)$ are within $\epsilon$ for every $\mu, \mu^\prime \in \Delta(\Theta)$, uniformly across all $s^{n}$.

For any $\epsilon> 0$, pick such $n$. Take any extreme point $\mu$ of $\eta_{\rho, \Pi}^\infty$. By \autoref{lem:D1}, we can find $s^n$ and $\hat \mu \in \eta_{\rho, \Pi}^\infty$ such that $\mu = r_{\rho, \Pi}(s^n|\hat \mu)$. By the above property, $r_{\rho, \Pi}(s^n|\mu^\prime)$ is within $\epsilon$ of $\mu$, independent of the initial belief $\mu^\prime \in \Delta(\Theta)$.
\end{proof}

Let $Q^{\Pi}(E)$ be the convex hull of the posterior beliefs that can arise with $\Pi$ when the prior belief lies in $E$.\footnote{$Q^{\Pi}(E) = co\left(\bigcup_{\mu \in E} \supp(Q_\mu^\Pi)\right)$.} The next lemma shows that, for any compact set $E$ of full-support beliefs, $Q^{\Pi}(E)$ is in the relative interior of $Q^{\Pi'}(E)$, and $\eta_{\rho, \Pi}(E)$ is in the relative interior of $\eta_{\rho, \Pi'}(E)$, when $\Pi$ is a weighted garbling of $\Pi'$ for a regular pair of $\Pi$ and $\Pi'$.

\begin{lem}
\label{lem:D3}
Suppose that $\Pi$ and $\Pi'$ is a regular pair of finite experiments and $\Pi$ is a weighted garbling of $\Pi'$. For any compact set $E$ of full-support distributions in $\Delta(\Theta)$, $Q^{\Pi}(E)$ is in the relative interior of $Q^{\Pi'}(E)$, and $\eta_{\rho, \Pi} (E)$ is in the relative interior of $\eta_{\rho, \Pi'} (E)$ for any $\rho \in R$.
\end{lem}

\begin{proof}
Take any compact set $E$ of full-support distributions. For any extreme point $\tilde \mu$ of $Q^{\Pi}(E)$, there is $\mu \in E$ and $s \in S$ such that $\tilde \mu$ is in the support of $Q^{\Pi}_{\mu}$. By regularity, the support of $Q^{\Pi}_{\mu}$ is in the relative interior of $Q^{\Pi'}({\mu}))$. Since $Q^{\Pi'}({\mu}) \subset Q^{\Pi'}(E)$, $\tilde \mu$ is in the relative interior of $Q^{\Pi'}(E)$. Since $Q^{\Pi}(E)$ is a compact convex set, it is the convex hull of all such extreme points. Hence $Q^{\Pi}(E)$ is in the relative interior of $Q^{\Pi'}(E)$.

Take any $\rho \in R$. Note that, for any compact set $E$ of full-support distributions, $\rho E = \{\rho \mu| \mu \in E\}$ is a compact set of full-support distributions. Hence it follows from the above result that $Q^{\Pi}(\rho E)$ is in the relative interor of $Q^{\Pi'}(\rho E)$. This immediately implies the result because $Q^{\Pi}(\rho E) = \eta_{\rho, \Pi} (E)$ and $Q^{\Pi'}(\rho E) = \eta_{\rho, \Pi'} (E)$.
\end{proof}

Now we prove \autoref{thm:4}.

\begin{proof}

First, we define some function given a pair of nonempty compact sets that are nested. Let $C$ and $D$ be any nonempty compact sets in $\mathbb{R}^n$ such that $C \subseteq D$ and $D$ is not a singleton. 
Let $\Lambda_D$ be the set of the unit-length vectors that are parallel to the affine hull of $D$ (i.e., the intersection of  $\{\lambda \in \mathbb{R}^n: \|\lambda\| =1 \}$ and the linear subspace that is a translate of the affine hull of $D$. Note that this set is well defined since the affine hull is not a point.) For each $\lambda \in \Lambda_D$, let $\xi(\lambda; D/C) = \max_{\mu \in D} \lambda \cdot \mu - \max_{\mu \in C} \lambda \cdot \mu$, which is nonnegative and strictly positive when $C$ is in the relative interior of $D$. Then let $\underline{\xi}(D/C) = \min_{\lambda \in \Lambda_D} \xi(\lambda; D/C)$. This is strictly positive when $C$ is in the relative interior of $D$ by the compactness of $\Lambda_D, C,$ and $D$.

Suppose that $\Pi$ and $\Pi'$ is a regular pair of finite experiments and $\Pi$ is a weighted garbling of $\Pi'$. Fix any $\rho \in R$. Let $E'$ be the set of initial belief $\mu_0$ that satisfy the assumption $\supp(Q^{\Pi'}_{\mu_0}) \subset \eta^\infty_{\rho, \Pi'}$, which is a compact set of full-support distributions. Then $Q^{\Pi'}(E') \subset \eta^\infty_{\rho, \Pi'}$ by the convexity of $\eta^\infty_{\rho, \Pi'}$. Since $Q^{\Pi}(E')$ is in the relative interior of $Q^{\Pi'}(E')$ by \autoref{lem:D3}, $Q^{\Pi}(E')$ is in the relative interior of $\eta^\infty_{\rho, \Pi'}$. Since $\eta^\infty_{\rho, \Pi'}$ is a compact convex set of full-support distribitions, $\eta_{\rho, \Pi}(\eta^\infty_{\rho, \Pi'})$ is in the relative interior of $\eta_{\rho, \Pi'}(\eta^\infty_{\rho, \Pi'})$ by \autoref{lem:D3}. Recall that $\eta_{\rho, \Pi'}(\eta^\infty_{\rho, \Pi'}) = \eta^\infty_{\rho, \Pi'}$ by \autoref{lem:D1}. Now let $C$ be the convex hull of the union of $Q^{\Pi}(E')$ and $\eta_{\rho, \Pi}(\eta^\infty_{\rho, \Pi'})$. Then $C$ is a compact convex set in the relative interior of $\eta^\infty_{\rho, \Pi'}$. Let $\epsilon =\underline{\xi}(\eta^\infty_{\rho, \Pi'}/C) > 0$. 

For any $\mu_0 \in E'$, the posterior belief in period $0$ for $\Pi$ is in $Q^{\Pi}(E')$, hence in $C$. Note that $\eta_{\rho, \Pi}^t(\eta^\infty_{\rho, \Pi'})$ is in $C$ for any $t$ since $\eta_{\rho, \Pi}(\eta^\infty_{\rho, \Pi'}) \subset \eta_{\rho, \Pi'}(\eta^\infty_{\rho, \Pi'}) = \eta^\infty_{\rho, \Pi'}$ and $\eta_{\rho, \Pi}(\cdot)$ is monotone with respect to belief set. This implies that the posterior beliefs always stay in $C$ with probability $1$ for $\Pi$ when the initial belief $\mu_0$ is in $E'$. Also note that the posterior beliefs always stay in $\eta^\infty_{\rho, \Pi'}$ with probability $1$ for $\Pi'$.

 Consider any decision problem $\mathcal{A} =(A, u, \mu_0) \in \mathcal{F}(\rho, \Pi')$. Let $\mu^*$ and $a^*$ be the solution for $\max_{\mu \in C} \max_{a \in A} u(a) \cdot \mu$, where $u(a) = u(a, \cdot)$ is the $|\Theta|$-dimensional payoff vector given $a \in A$. Since $\mu_0 \in E'$, the posterior belief for $\Pi$ always stays in $C$, hence $V^{\mathcal{A}, \rho, T}(\Pi)$ is bounded above by $u(a^*) \cdot \mu^*$. Let $\mu'$ be an extreme point of $\eta^\infty_{\rho, \Pi'}$ that solves $\max_{\mu \in \eta^\infty_{\rho, \Pi'}} u(a^*) \cdot \mu$ and $\underline{\mu}$ be a solution for $\min_{\mu \in \eta^\infty_{\rho, \Pi'}} u(a^*) \cdot \mu$.

We consider two cases. First suppose that $u(a^*)$ is $0$ vector or orthogonal to the affine hull of $\eta^\infty_{\rho, \Pi'}$. This is the case where the DM's payoff $u(a^*) \cdot \mu$ is constant on $\eta^\infty_{\rho, \Pi'}$ (hence on $C$), which corresponds to the optimal payoff given $\Pi$. In this case, we immediately obtain $V^{\mathcal{A}, \rho, T}(\Pi') \geq V^{\mathcal{A}, \rho, T}(\Pi)$ for any $T$.

Next, suppose that $u(a^*)$ is a nonzero vector that is not orthogonal to the affine hull of $\eta^\infty_{\rho, \Pi'}$. Let $\hat u(a^*)$ be the projection of $u(a^*)$ to the linear subspace that is a translate of the affine hull. Note that $\frac{\hat u(a^*)}{\|\hat u(a^*)\|} (\mu' - \mu^*) > \epsilon$ independent of $\mathcal{A} \in \mathcal{F}(\rho, \Pi')$, since $\frac{\hat u(a^*)}{\|\hat u(a^*)\|} \in \Lambda_{\eta^\infty_{\rho, \Pi'}}$. 

Pick $\epsilon' \in (0,1)$ that satisfies $(1- \epsilon') (\epsilon - \epsilon') - \sqrt{2} \epsilon'  > 0$. By \autoref{lem:D2}, we can pick $T'$ (independent of $\mathcal{A}$) such that every extreme point of $\eta^\infty_{\rho, \Pi'}$ can be approximated within $\epsilon'$ at least once with probability $1 - \epsilon'$ within $T$ periods for any $T\geq T'$.

Consider the following strategy with $\Pi'$ for problem $(\mathcal{A}, \rho, T)$: choose $a^*$ as soon as the posterior belief is within $\epsilon'$ distance from $\mu'$ or on the terminal date $T$ otherwise. If the DM chooses $a^*$ before the end of the game, the DM achieves $u(a^*) \cdot \mu$ for some posterior belief $\mu$ that is within $\epsilon'$ distance from $\mu'$. Let $\tilde \mu$ be the expected value of this posterior belief conditional on this event, which is within $\epsilon'$ from $\mu'$. The value of this strategy is at least $(1-\epsilon') u(a^*) \cdot  \tilde \mu + \epsilon' u(a^*) \cdot \underline{\mu}$, since the posterior belief is always in $\eta^\infty_{\rho, \Pi'}$ and $\underline{\mu}$ is the worst posterior belief in $\eta^\infty_{\rho, \Pi'}$ to play $a^*$.

Hence, for any  $\mathcal{A} \in F(\rho, \Pi')$ and $T \geq T'$, 
\begin{align*}
V^{\mathcal{A}, \rho, T}(\Pi') -V^{\mathcal{A}, \rho, T}(\Pi) &\geq (1- \epsilon') u(a^*) \cdot \tilde \mu + \epsilon' u(a^*) \cdot \underline{\mu} - u(a^*) \cdot \mu^* \\
& = (1- \epsilon') u(a^*) \cdot (\tilde \mu - \mu^*) - \epsilon' u(a^*) \cdot (\mu^*  - \underline{\mu}) \\
& =(1- \epsilon') \hat u(a^*) \cdot (\tilde \mu - \mu^*) - \epsilon' \hat u(a^*) \cdot (\mu^*  - \underline{\mu}) \\
& = \|\hat u(a^*)\| \left[(1- \epsilon') \frac{\hat u(a^*)}{\|\hat u(a^*)\|} \cdot (\tilde \mu - \mu^*) - \epsilon' \frac{\hat u(a^*)}{\|\hat u(a^*)\|} \cdot (\mu^*  - \underline{\mu}) \right] \\
& \geq  \|\hat u(a^*)\| \left[(1- \epsilon') \left(\frac{\hat u(a^*)}{\|\hat u(a^*)\|} \cdot (\mu' - \mu^*) -\frac{\hat u(a^*)}{\|\hat u(a^*)\|} \cdot (\mu' - \tilde \mu)\right) -  \sqrt{2} \epsilon' \right] \\
& \geq  \|\hat u(a^*)\| \left[(1- \epsilon') (\epsilon - \epsilon') - \sqrt{2} \epsilon'  \right] \\
& \geq 0,
\end{align*}
where the inequalities follow from $\frac{\hat u(a^*)}{\|\hat u(a^*)\|} \cdot (\mu' - \mu^*) \geq \underline{\xi}(\eta^\infty_{\rho, \Pi'}/C) = \epsilon$ and the Cauchy-Schwarz inequality.

\vspace{2mm}

Conversely, suppose that $\Pi$ is not a weighted garbling of $\Pi^\prime$. We show that there is a $\tilde \rho \in R$ and $\mathcal{\tilde A} \in \mathcal{F} (\tilde \rho, \Pi')$ such that there does not exist such a lower bound $T'$.

Take any full-support distribution $\tilde \mu \in \Delta(\Theta)$. By \autoref{thm:2} there exists some posterior belief $\hat \mu$ in the support of $Q^\Pi_{\tilde \mu}$ that is outside of $Q^{\Pi'}(\{\tilde \mu \})$.
Hence there exists $\mathcal{\tilde A} =(A, u, \tilde \mu)$ such that $u(a) \cdot \hat \mu > \max_{\mu \in Q^{\Pi'}(\{\tilde \mu \})} \max_{a \in A} u(a) \cdot \mu$ for some $a \in A$.

Consider the i.i.d.process $\rho^{\tilde \mu} \in R$. Note that $\mathcal{\tilde A} \in \mathcal{F}(\rho^{\tilde \mu}, \Pi')$ since the initial belief is $\tilde \mu$ for $\mathcal{\tilde A}$. Since the posterior belief always stays in $\eta^{\infty}_{\rho^{\tilde \mu}, \Pi'} = Q^{\Pi'}(\{\tilde \mu \})$ for $\Pi'$, $V^{\mathcal{\tilde A}, \rho^{\tilde \mu}, T}(\Pi^\prime)$ is bounded above by $\max_{\mu \in Q^{\Pi'}(\{\tilde \mu \})} \max_{a \in A} u(a) \cdot \mu$. On the other hand, $\lim_{T \rightarrow \infty} V^{\mathcal{\tilde A}, \rho^{\tilde \mu}, T}(\Pi) = u(a) \cdot \hat \mu$ since the belief $\hat \mu$ realizes almost surely in the limit as $T \rightarrow \infty$ for $\Pi$. Hence, there cannot exist $T'$ that satisfies the condition in the second item. 
\end{proof}

\subsection{Proof of \autoref{prop:n3}}

\textbf{Proof of Item 1:}
We show that there are experiments $\Pi$ and $\Pi'$ such that $\Pi' \succeq_{LD} \Pi$, while $\Pi \not \succeq_{WG}\Pi'$.	

Given two experiments $\Pi_A$, $\Pi_B$ and $\eta \in [0,1]$, let $\eta \Pi_A \oplus (1-\eta) \Pi_B$ be the experiment in which the signal is generated by $\Pi_A$ with probability $\eta$ and by $\Pi_B$ with probability $1-\eta$. Let $\varnothing$ denote the null experiment. 
 
 Let $\Pi$ and $\Pi'$ be the experiments that satisfy the conditions in \autoref{prop:pp3}. We claim that there exists $\tilde{\Pi}  := \eta \Pi \oplus (1-\eta) \varnothing$ for some $\eta \in [0,1]$ (to be chosen below) such that 
\begin{equation}
\label{eq:001}
  \Pi' \not  \succeq_{WG}   \tilde{\Pi} , \  \Pi^{ \prime  \otimes 2} \succeq_{BG} \tilde{\Pi}^{ \otimes 2}, \ \Pi^{ \prime  \otimes 3} \succeq_{BG} \tilde{\Pi}^{ \otimes 3},	
\end{equation}
where $\succeq_{BG}$ represents the Blackwell order.

We first observe the following.

\begin{lem}
\label{lem:oo2}
For each $n \in \mathbb{N}$, if $\eta' \geq 1- (1-\eta)^n$, $ \eta' \Pi^{ \otimes n}  \oplus (1-\eta') \varnothing  \succeq_{BG} ( \eta \Pi \oplus  (1- \eta) \varnothing )^{ \otimes n}$. 
\end{lem}	
	
	\begin{proof}
Note that $\Pi^{ \otimes n} \succeq_{BG} (\Pi^{ \otimes k} \otimes \varnothing^{\otimes n-k})$ for any $1 \leq k \leq n$. Under $( \eta \Pi \oplus  (1- \eta) \varnothing )^{ \otimes n}$, the probability that the signal is generated from $\Pi$ at least once is $ 1- (1-\eta)^n$. Hence, if $\eta' \geq 1- (1-\eta)^n$, $ \eta' \Pi^{ \otimes n}  \oplus (1-\eta') \varnothing \succeq_{BG} ( \eta \Pi \oplus  (1- \eta) \varnothing )^{ \otimes n}$. 
\end{proof}

Since $\Pi^{ \prime \otimes 2} \succeq_{WG} \Pi^{ \otimes 2}$, \autoref{thm:n1} implies that there exists $\eta' \in (0,1]$ such that 
$\Pi^{ \prime \otimes 2} \succeq_{BG} \eta' \Pi^{ \otimes 2} \oplus (1-\eta') \varnothing$. Then \autoref{lem:oo2} implies that there exists $\eta_2 \in (0,\eta']$ sufficiently small such that $ \eta' \Pi^{ \otimes 2} \oplus (1-\eta') \varnothing \succeq_{BG} ( \eta_2 \Pi \oplus  (1- \eta_2) \varnothing )^{ \otimes 2 }$. Then by transitivity of the Blackwell order, it follows that $\Pi^{ \prime \otimes 2} \succeq_{BG} ( \eta_2 \Pi \oplus  (1- \eta_2) \varnothing )^{ \otimes 2}$. A similar argument yields $\eta_3 \in (0,\eta']$ such that $\Pi^{ \prime \otimes 3} \succeq_{BG} ( \eta_3 \Pi \oplus  (1- \eta_3) \varnothing )^{ \otimes 3}$. Let $\eta : = \min \{ \eta_2, \eta_3\} >0$, and define $\tilde{\Pi}: = \eta \Pi  \oplus (1-\eta) \varnothing $. Since $\tilde{\Pi}$ shares the same extreme points of the induced posterior beliefs, $\Pi' \not \succeq_{WG} \tilde{\Pi}$, so we obtain \eqref{eq:001}.

We use the following result from \cite{Blackwell_1951} (Theorem 12).
\begin{lem}
\label{lem:oo3}
Let $\Pi_k$ be an experiment for $k=1,2,3,4$. If $\Pi_1  \succeq_{BG} \Pi_2 $ and $\Pi_3 \succeq_{BG} \Pi_4$, $\Pi_1 \otimes \Pi_3 \succeq_{BG} \Pi_2 \otimes \Pi_4$.
\end{lem}

By \autoref{lem:oo3}, $\Pi^{ \prime  \otimes 2} \succeq_{BG} \tilde{\Pi}^{ \otimes 2}$ implies that any even-number samples from $\Pi'$ is better than the same number of samples from $\tilde{\Pi}$. Then note that any odd number $m= 2k + 3$ for some $k \in \{ 0,1,2,\dots\}$, i.e., an even number plus three. Since $\Pi^{ \prime  \otimes 3} \succeq_{BG} \tilde{\Pi}^{ \otimes 3}$, it follows that for any $n \geq 2$, $\Pi^{\prime \otimes n}$ is better than $\tilde{\Pi}^{ \otimes n}$. Hence, $\Pi' \succeq_{LD} \tilde{\Pi}$. However, as noted above, $\Pi' \not \succeq_{WG} \tilde{\Pi}$.

\vspace{0.5cm}

Finally consider the two-state case with $\Theta = \{0,1\}$. If $\Pi$ generates a more extreme likelihood ratio than $\Pi'$ (to be precise, if $ess\sup_{s \in S} \frac{d\pi_\theta(s)}{d\pi_{1-\theta}(s)} > ess\sup_{s' \in S'} \frac{d\pi'_\theta(s')}{d\pi'_{1-\theta}(s')}$ for $\theta =0$ or $ 1$), then $\Pi' \not \succeq_{LD} \Pi$ since $\Pi'$ does not dominate $\Pi$ in the R\'{e}nyi order, which is a necessary condition for the large sample dominance order (See Section 5.1 of \cite{MPST_2021_ECMA}). Therefore, if $\Pi' \succeq_{LD} \Pi$, then $ess\sup_{s' \in S'} \frac{d\pi'_\theta(s')}{d\pi'_{1-\theta}(s')} \geq ess\sup_{s \in S} \frac{d\pi_\theta(s)}{d\pi_{1-\theta}(s)} $ for $\theta =0$ and $1$. For finite experiments, this means that the maximum likelihood ratios given $\theta = 0, 1$ for $\Pi'$ (i.e. $\max_{s'}\frac{\pi_\theta'(s')}{\pi'_{1-\theta}(s')}$) are at least as large as those for $\Pi$. Hence $\Pi' \succeq_{WG}\Pi $ by \autoref{thm:2}.

\vspace{0.5in} 

\textbf{Proof of Item 2:} We reproduce an example from \cite{Azrieli_2014_ECMA}:
\begin{figure}[h]
\centering
\begin{tabular}{|c|c|c|}
\hline
$\pi_\theta$    & $\theta_1$  & $\theta_2$  \\ \hline
$s_1$ & $\alpha$    & $1- \alpha$ \\ \hline
$s_2$ & $1- \alpha$ & $\alpha$    \\ \hline
\end{tabular}
\quad 
\begin{tabular}{|c|c|c|}
\hline
$\pi'_\theta$    & $\theta_1$  & $\theta_2$  \\ \hline
$x_1$ & $\beta$     & $\frac{1}{2}-\beta$ \\ \hline
$x_2$ & $\frac{1}{2}$       & $\frac{1}{2}$       \\ \hline
$x_3$ & $\frac{1}{2}-\beta$ & $\beta$     \\ \hline
\end{tabular}
\caption{}
\label{fig:3}
\end{figure}
Consider the experiments in \Autoref{fig:3}, denoted by $\Pi(\alpha)$ and $\Pi'(\beta)$, respectively, where $\alpha \in [0,\frac{1}{2} ]$ and $\beta \in [0,\frac{1}{4} ]$. Note that under $\Pi(\alpha)$, signal $s_i$ suggests that the state is $\theta_j$ for $i=1,2$ and $j \neq i$. A similar remark applies to $x_1$ and $x_3$ Under $\Pi' (\beta)$. We observe:
\begin{enumerate}
	\item If $\alpha \geq 2\beta$, then $\Pi' (\beta) \succeq_{WG} \Pi (\alpha)$.
	\item \cite{Azrieli_2014_ECMA} shows that if $\alpha < h(\beta) = \frac{1}{2} ( 1- \sqrt{\frac{3}{4}  + 4\beta^2 - 2 \beta -2 \sqrt{\beta ( \frac{1}{2} - \beta)}} )$, then $\Pi (\alpha) \succeq_{MS} \Pi' (\beta)$ (and $\Pi' (\beta) \not \succeq_{MS} \Pi (\alpha))$), whereas $\Pi(\alpha) \not \succeq_{LD} \Pi' (\beta)$. On the other hand, if $\alpha > h(\beta)$, $\Pi' (\beta) \succeq_{MS} \Pi(\alpha)$.
	\item $h(\beta) > 2\beta$ for any $\beta \in [0,\frac{1}{4})$ and $h(\beta) = 2 \beta$ when $\beta = \frac{1}{4}$.
\end{enumerate}

Hence, for $\alpha$ and $\beta$ such that $2\beta < \alpha <h(\beta)$, $\Pi' (\beta) \succeq_{WG} \Pi (\alpha)$, but $\Pi' (\beta) \not \succeq_{MS} \Pi (\alpha)$, hence $\Pi' (\beta) \not \succeq_{LD} \Pi (\alpha)$.